\documentclass[11pt,a4paper]{article}
\usepackage{jheppub}
\usepackage[utf8]{inputenc}
\usepackage{graphicx,fancybox,float,comment}
\usepackage{pdflscape}
\usepackage{units}
\usepackage{multirow,array,arydshln}
\usepackage[dvipsnames]{xcolor}
\usepackage{braket,tensor}
\usepackage{amsmath,amssymb,amsfonts,mathrsfs}
\usepackage{tikz-cd}
\usepackage{mdframed}
\usepackage{subfig}
\usepackage[notref,notcite]{}
\usepackage{bm}

\usepackage{ulem}

\global\interfootnotelinepenalty=1000

\graphicspath{{graphics}}

\newcommand{\bea}{\begin{eqnarray}}
\newcommand{\eea}{\end{eqnarray}}
\newcommand{\be}{\begin{equation}}
\newcommand{\ee}{\end{equation}}

\newcommand{\hp}[1]{\hphantom{#1}}

\def\({\left(}
\def\){\right)}

\def \a {\alpha}
\def \b {\beta}

\def \d {\delta}
\def \e {\epsilon}
\def \g {\gamma}
\def \k {\kappa}
\def \o {\omega}

\def \r {\rho}
\def \l {\lambda}
\def \m {\mu}
\def \n {\nu}
\def \s {\sigma}

\def \th {\theta}

\def \D {\Delta}
\def \G {\Gamma}
\def\pa{\partial}
\def\nn {\nonumber}
\def\ex{\mathrm{e}}
\def\OO{\mathcal{O}}
\def\GG{\mathcal{G}}
\def\AA{\mathcal{A}}
\def\intd{\mathrm{d}}

\subheader{\begin{flushright}
\end{flushright}}

\title{Near-extremal hydrodynamics and the holographic product formula}

\author[]{Edwan Pr\'eau}
\affiliation[]{Institute for Theoretical Physics, Utrecht University, 3584 CC Utrecht, The Netherlands\\}
\emailAdd{e.c.m.preau@uu.nl}

\abstract{The holographic product formula is used to determine the general form taken by holographic spectral functions in the near-extremal hydrodynamic regime, with energy $\o$, momentum $k$ and temperature $T$ much smaller than a hard scale $\m$. The resulting expressions simplify in the extremal limit $T \ll \o,k\ll \m$, for which the low-temperature gapless modes and the IR conformal behavior factorize. In some cases, this factorization extends to the general near-extremal regime $\o,k,T\ll\m$ at leading order in $T/\m$. Several examples are discussed with different types of gapless modes and IR CFTs, including new numerical results for low temperature quasi-normal modes. We end with a concrete application that shows how the inclusion of the IR conformal behavior improves the description of the spectral function at low energies. }

\begin{document}
\maketitle
\flushbottom

\section{Introduction and summary}
\label{sec::intro}

Most finite temperature field theories are described by hydrodynamics over long times and distances. That is, for energies $\o$ and momenta $k$ much smaller than temperature $T$, the dynamics of most systems reduces to the conservation equations for their conserved currents. This includes in particular the majority of holographic theories, whose gravitational dual should therefore also admit a fluid description in the hydrodynamic regime. This result was established explicitly in \cite{Bhattacharyya:2007vjd}, and has been called the \textit{fluid/gravity correspondence}.

A more surprising result that has since been observed in holographic systems at finite density, is that hydrodynamics seemingly continues to hold in the low temperature limit $T\ll\m$ (with $\m$ the chemical potential), as long as energies and momenta remain much smaller than $\m$ \cite{Edalati:2010hk,Edalati:2010pn,Moitra:2020dal,Davison:2013bxa,Davison:2022vqh,Jarvinen:2023xrx,Arnaudo:2024sen}. In particular, at leading order in the low $\o,k$ expansion, holographic two-point functions were found to agree with the zero-temperature limit of hydrodynamic correlators \cite{Davison:2013bxa,Moitra:2020dal}. Not only do the hydrodynamic poles survive, but no sign was found at this order of the branch cut that is known to arise at $T=0$ \cite{Faulkner:2009wj,DHoker:2010xwl}, which is related to the lower-dimensional conformal field theory (CFT) emerging in the infra-red (IR) of these holographic systems.  

The recent calculation of \cite{Gouteraux:2025kta} at next-to-leading order made it clear that the branch cut puzzle only exists at leading order, since the next order already contains logarithmic signatures of the branch cut. However, the result of \cite{Gouteraux:2025kta} still does not contain a clear imprint from the IR CFT.

\begin{figure}[h]
\begin{center}
\includegraphics[scale=0.45]{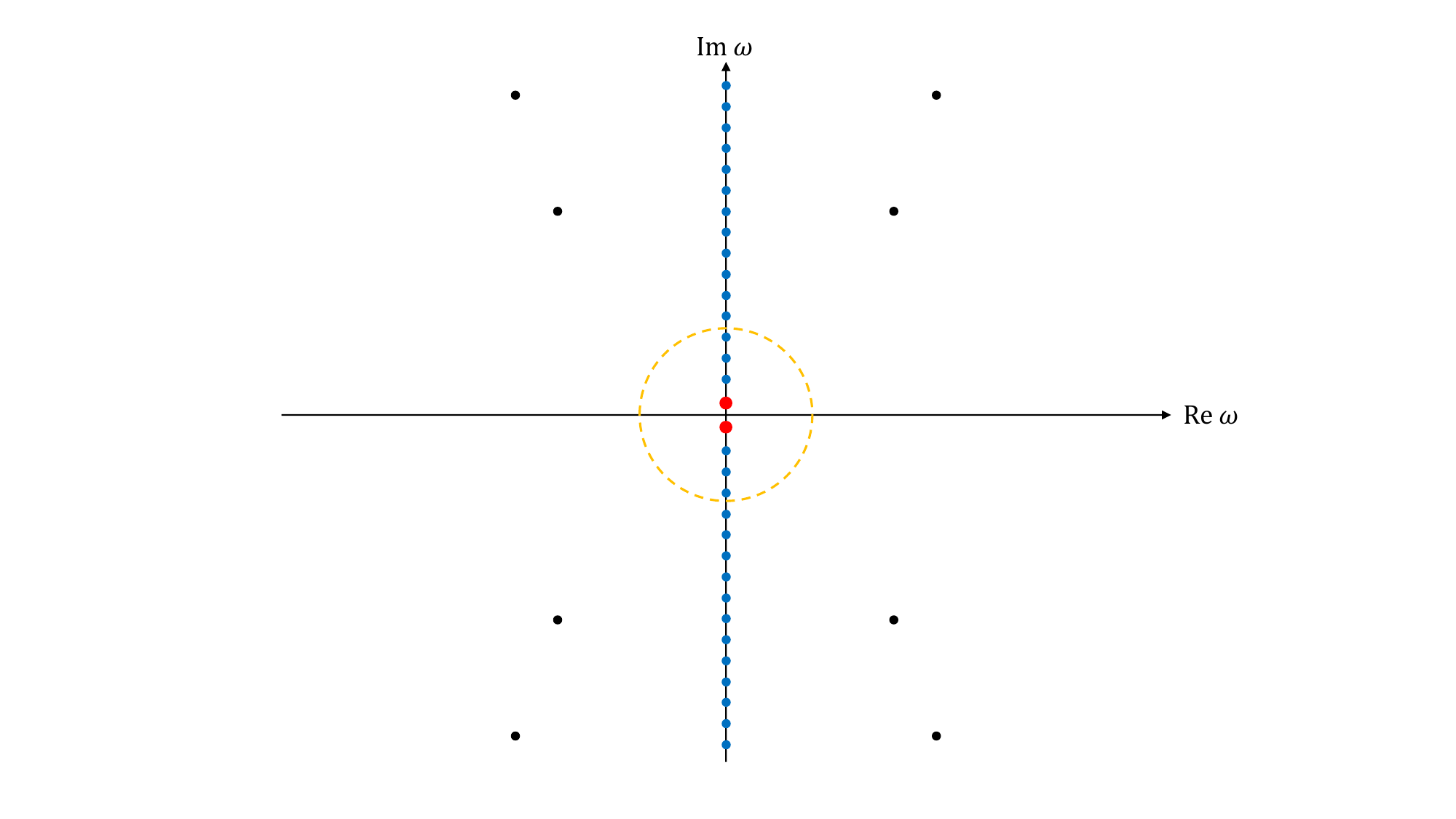}
\caption{Schematic pole structure of holographic spectral functions in the near-extremal regime $T\ll\m$. The black dots indicate hard poles of order $\OO(\m)$, the blue dots soft poles of order $\OO(T)$ and the red dot is an example of gapless pole. The near-extremal hydrodynamic region (bounded by the yellow-dashed line) contains only soft and gapless poles. This picture is close to the actual pole structure of the correlators in section \ref{sec:sound}, but more general behaviors are possible. In particular the soft poles of section \ref{sec:mb} have finite real parts.}
\label{Fig:NEpoles}
\end{center}
\end{figure}

In this work, we exploit the special analytic properties of holographic correlators to get more insight into their structure in the near-extremal hydrodynamic regime $\o,k,T\ll\m$. Specifically, we derive the consequences of the recently established holographic product formula \cite{Dodelson:2023vrw}, which makes it possible to express holographic spectral functions as a product over their poles\footnote{To be more precise, the result of \cite{Dodelson:2023vrw} requires some conditions on the correlators of interest, which are not common to all holographic correlators. These conditions are reviewed in section \ref{sec:PF}. As mentioned in section \ref{discussion}, it would be interesting to investigate how much these conditions can be extended.}. This formula therefore allows to write the general form of near-extremal holographic correlators based on their known pole structure, which features three types of poles (see figure \ref{Fig:NEpoles}): the hard poles with gaps of order $\OO(\m)$, the soft poles with gaps of order $\OO(T)$ (related to the IR CFT), and the gapless poles, which behave like hydrodynamic modes at leading order in the near-extremal hydrodynamic limit. From this structure, we deduce the \textit{near-extremal hydrodynamic product formula}
\begin{equation}
\label{I1} \rho(\o,k)  = \sinh\left(\frac{\o}{2T}\right)\frac{\g(\o,k)\GG_s(\o,k)}{\prod_{n\in\text{gapless}}(\o^2-\o_n^2)(\o^2-(\o_n^*)^2)} \, ,
\end{equation}
with $\r$ the spectral function, $\GG_s$ the inverse product over the soft poles and $\g$ a generalized susceptibility, which is analytic for $\o\ll\m$. 

Due to the close relation between the soft poles and the poles of the IR CFT correlator, it is expected that $\GG_s$ should also somehow be related to the IR correlator $\GG_{\text{IR}}$. However, it is not true that $\GG_s$ and $\GG_{\text{IR}}$ are equal in the general near-extremal hydrodynamic regime, which makes the above product formula of limited use in general. Instead, the product formula is most useful in the limit of vanishing temperature -- i.e. for $T\ll\o,k\ll\m$ -- where the soft poles merge into a branch cut, and the low $\o$ behavior of $\r$ can be shown to be controlled by the IR CFT$_q$ : $\r\sim(\o^2-\vec{k}_{q-1}^2)^{\D(k)-q/2}$, with $\D(k)$ the momentum-dependent IR conformal dimension and $\vec{k}_{q-1}$ the spatial momentum along the IR CFT directions. From this, we infer the \textit{extremal hydrodynamic product formula}
\begin{equation}
\label{I2} \r  = \frac{\g(\o,k)(\o^2-\vec{k}_{q-1}^2)^{\D(k)-q/2}}{\prod_{n\in\text{gapless}}(\o^2-\o_n^2)(\o^2-(\o_n^*)^2)} \, ,
\end{equation}
where $\g$ is finite at $\o=|\vec{k}_{q-1}|$, and is expected to follow a generalized (non-local) hydrodynamic expansion for $\o,k \ll \m$, including a mixture of polynomial and logarithmic terms. 

In some cases, the soft poles can be shown (numerically) to precisely approach the IR poles in the near-extremal limit, up to corrections of order $\OO(T/\m)$ \cite{Edalati:2010hk,Davison:2022vqh}. In these cases, the near-extremal product formula simplifies at leading order in $T/\m$ to
\begin{equation}
\label{I3} \r = \sinh\left(\frac{\o}{2T}\right)\frac{\tilde{\g}(\o,k)\GG_{\text{IR}}(\o,k)}{\prod_{n\in\text{gapless}}(\o^2-\o_n^2)(\o^2-(\o_n^*)^2)}(1+\OO(T/\m)) \, ,
\end{equation}
with $\GG_{\text{IR}}$ the thermal correlator in the IR CFT, and $\tilde{\g}$ analytic. This expression is valid in the general near-extremal hydrodynamic regime, including when $\o$ is of the same order as the temperature.

The last two product formulae \eqref{I2} and \eqref{I3} are the main results emphasized in this work. Whenever they apply, these formulae imply a simple factorized structure for holographic spectral functions in the (near-)extremal hydrodynamic limit, where both the low-temperature gapless modes and the IR conformal behavior appear explicitly. 

We would like to emphasize the universal nature of the product formulae \eqref{I2} and \eqref{I3}, for all systems in the class studied in this work, that have an emergent conformal description in the IR. In particular, these formulae do not require to compute explicitly the soft poles\footnote{
In the case of \eqref{I3}, it is in fact necessary to compute (in general numerically) the soft modes, to verify whether the poles have the required property of being parametrically close to the IR poles. However, once the formula \eqref{I3} is motivated by some calculation of the soft poles, it provides an analytic approximation to the correlator which does not require the precise data for these poles, since they are approximated by the exact IR poles. 
Note that the examples discussed in this work (sections \ref{sec:ccc} and \ref{sec:mb}) and in the literature indicate that \eqref{I3} applies in the absence of gapless poles. In presence of gapless poles, we found that \eqref{I3} may apply (section \ref{sec:sound}) or not (section \ref{sec:ccc}).}, and express instead the spectral function in terms of three quantities: the infra-red scaling dimension $\D(k)$, the gapless-poles $\omega^g_n(k)$ and the generalized susceptibility $\gamma$ (or $\tilde{\gamma}$). $\D(k)$ can be computed from the standard near-horizon limit of \cite{Faulkner:2009wj} (see appendix \ref{sec::AppA}). The gapless poles and susceptibility can also be computed analytically, order by order in $\omega/\mu$ and $k/\mu$ \cite{Davison:2013bxa,Gouteraux:2025kta} (although the calculation becomes rather involved beyond leading order \cite{Gouteraux:2025kta}).

The rest of this paper is organized as follows. Section \ref{sec:PF} discusses in more details the developments summarized above, whereas section \ref{sec:Ex} presents several examples where the product formula can be used to determine the general structure of near-extremal spectral functions. This includes probe gauge field (section \ref{sec:ccc}) and helicity-0\footnote{The helicity-1 shear-channel of metric perturbations has been analyzed in the recent work \cite{Bhattacharya:2025vyi}.} metric (section \ref{sec:sound}) perturbations on the AdS$_5$-Reissner-Nordstr\"om background -- with a (0+1)-dimensional CFT in the IR -- as well as scalar perturbations of magnetized black branes (section \ref{sec:mb}), whose IR description is in terms of a (1+1)-dimensional CFT \cite{DHoker:2009mmn,DHoker:2010xwl}. As intermediate results, we computed numerically the low-lying poles of the spectral functions in the near-extremal regime for all three cases. These numerical results are presented in appendices \ref{sec::AppB3}, \ref{Sec::AppE} and \ref{Sec::AppH}. 

Section \ref{sec:application} discusses a concrete application of the product formula to the problem analyzed in \cite{Jarvinen:2023xrx}, related to neutrino transport in dense holographic matter. Improving the hydrodynamic approximation to the weak-current correlators with the IR scaling is shown to give a better description of the neutrino opacity near the Fermi surface (see figure \ref{Fig:kaenu}). 

Several additional details are provided in the appendices. In particular, in appendix \ref{sec::AppB}, we show how the next-to-leading order result of \cite{Gouteraux:2025kta} is consistent with the product formula, which completes it at low energy $\o$ by having the IR power-law appear explicitly.   

\section{The product formula}

\label{sec:PF}

We present in this section the product formula for holographic two-point functions, as introduced in \cite{Dodelson:2023vrw}. We first review the formula and its conditions of application, and then work out its consequences for the structure of holographic correlators in the near-extremal regime.

\subsection{The formula}

The product formula \cite{Dodelson:2023vrw} is a statement about thermal two-point functions at equilibrium. For a given operator $\OO$, it is best expressed in terms of the two-sided correlator $G_{12}$, defined in position space as
\begin{equation}
\label{d1} G_{12}(t,\vec{x}) \equiv \frac{1}{Z(\beta)}\mathrm{Tr}\left[\ex^{-\beta H} \mathcal{O}\left(t-\frac{i\beta}{2},\vec{x}\right)\mathcal{O}(0,0) \right] \, , 
\end{equation}
with $\beta \equiv 1/T$ the inverse temperature, $H$ the Hamiltonian and $Z(\beta)\equiv \mathrm{Tr}[\ex^{-\beta H}]$ the partition function. $G_{12}$ corresponds to the Wightman function, but evaluated with an imaginary time shift $-i\b/2$. 

At equilibrium, the Wightman function is related to the retarded correlator 
\begin{equation}
\label{d4} G_R(t,\vec{x}) \equiv i\theta(t) \frac{1}{Z(\beta)}\mathrm{Tr}\left( \ex^{-\beta H}\left[\mathcal{O}(t,\vec{x}),\mathcal{O}(0,0)\right] \right) \, ,
\end{equation}
which may be written in Fourier space as 
\begin{equation}
\label{pf3} G_{12}(\omega,\vec{k}) = \frac{\mathrm{Im}G_R(\omega,\vec{k})}{\sinh{(\beta\omega/2)}} \equiv \frac{G_R(\omega,\vec{k}) - G_R(-\omega,\vec{k})}{2i\sinh{(\beta\omega/2)}} \, ,
\end{equation}
with $\omega$ the energy and $\vec{k}$ the spatial momentum\footnote{This relation holds when there is no chemical potential for the charges of the operator $\mathcal{O}$. Otherwise $\omega$ should be shifted by the appropriate chemical potentials $\mu$ in the argument of the sinh (rescaling also $G_{12}$ by a factor $\ex^{-\beta\mu/2}$).}. 

Now the product formula expresses certain holographic two-sided thermal correlators in Fourier space as a product over their poles $\omega_n(\vec{k})$\footnote{In the special case of purely imaginary poles, there is a single factor $(1-(\omega/\omega_n)^2)$.}
\begin{equation}
\label{pf1} G_{12}(\omega,\vec{k}) = \frac{\gamma(\vec{k})}{\prod_{n=1}^\infty\left(1-\frac{\omega^2}{\omega_n(\vec{k})^2}\right)\left(1-\frac{\omega^2}{(\omega_n^*(\vec{k}))^2}\right)} \, ,
\end{equation}
with $\gamma(\vec{k})$ free of zeroes and poles. Note in particular that the residues do not appear explicitly in the formula, and their data is fully captured by the poles and the function $\gamma(\vec{k})$. 

Equation \eqref{pf1} was derived in \cite{Dodelson:2023vrw} as a consequence of Hadamard's factorization theorem, based on the assumptions that $G_{12}(\omega)$ is meromorphic (its only singularities are isolated poles), has no zeroes, and behaves asymptotically as
\begin{equation}
\label{pf2} \left|G_{12}(\omega,\vec{k})\right| \underset{|\omega|\to\infty}{\sim} P(\omega) \ex^{-\l(\theta)|\omega|}, \quad\l(\th)\geq 0, \quad \omega\equiv |\omega|\ex^{i\th}   \, ,
\end{equation}
with $P(\omega)$ a polynomial. The previous conditions imply that $1/G_{12}(\omega)$ is an entire function of order 1 \cite{Dodelson:2023vrw}. Meromorphicity and the asymptotic behavior \eqref{pf2} are generic properties of holographic correlators \cite{Hartnoll:2005ju,Festuccia:2005pi,Festuccia:thesis}. Although the condition that $G_{12}$ be free of zeroes is a priori not generic, it was shown to hold in \cite{Dodelson:2023vrw} in the following conditions:
\begin{enumerate}
\item The dual fluctuation equations decouple, such that we can focus on a single equation dual to a given correlator. This equation can be written in the Schr\"odinger form\footnote{A given set of decoupled fluctuation equations can always be written in the Schr\"odinger form \eqref{pf2b} upon appropriate field redefinitions (see e.g. appendix B of \cite{Arean:2013tja}).} as
\begin{equation}
\label{pf2b} \psi''(z) + (\omega^2-V(z))\psi(z) = 0 \, ,
\end{equation}
with $z$ the tortoise coordinate, for which the background AdS black-hole metric takes the form
\begin{equation}
\label{pf2c} \intd s^2 = \ex^{2A(z)}(-f(z)\intd t^2 + dz^2+\intd s_{d-1}^2) \, ,
\end{equation}
with $A(z)$ the scale factor, $f(z)$ the blackening function and $\intd s_{d-1}^2$ the metric along the spatial directions of the $d$-dimensional boundary.

Note that finding variables that obey decoupled equations is not easy in general, and not always possible. A non-trivial example of such variables is given by the ``master fields" for the metric and gauge fields fluctuations of the RN background \cite{Kodama:2003kk,Edalati:2010pn,Edalati:2010hk}.

\item The Schr\"odinger potential $V(z)$ should be regular for all $z>0$, with boundary ($z\to 0$) and horizon ($z\to\infty$) asymptotics of the form
\begin{equation}
\label{pf2d} V(z) \underset{z\to 0}{\sim} \frac{\n^2-\frac{1}{4}}{z^2} \quad , \quad V(z) = \sum_{n\geq1} \ex^{-\frac{4\pi n}{\beta} z} \, ,
\end{equation}
with $\n^2\geq 0$. In practice, all regular holographic solutions obey the IR condition at finite temperature \cite{Dodelson:2023vrw}. The UV $z^{-2}$ behavior of the potential is also generic, although in some cases the parameter $\n^2$ may become negative. So the effective condition on the potential is the absence of singularities in the bulk -- i.e. for all $z\in (0,\infty)$ -- together with $\n^2\geq0$.
\end{enumerate}
In this work, we will also focus for simplicity on cases where the fluctuation equations decouple. As we mention later in the discussion, it would be interesting to investigate to what extent formulae of the form \eqref{pf1} still hold in the case of coupled equations.

On the other hand, it is most of the time not crucial that $G_{12}$ is actually free of zeroes, at least for a finite number of zeroes. If $G_{12}$ has a finite set of zeroes $\{\omega^z_m(k)\}_m$, it is indeed straightforward to generalize \eqref{pf1} since $G_{12}/\prod_m(\omega-\omega^z_m(k))$ is itself free of zeroes\footnote{The case of an infinite number of zeroes is more subtle, since the infinite product $\prod_m(\omega-\omega^z_m(k))$ may affect the asymptotic behavior \eqref{pf2}.}. Most of our results are then expected to go beyond the case of a regular Schr\"odinger potential, although it would be interesting to better understand in general the relation between singularities of the potential and zeroes of $G_{12}$.     

\subsection{Application to near-extremal hydrodynamics}

\label{sec:pfne}

We now consider a thermal state with a dimensionful source parameter $\m$, in the near-extremal limit $T \ll \m$, such that the bulk dual corresponds to a near-extremal black brane with AdS$_p$-Schwarzschild near-horizon geometry. The low energy excitations of such states are described by a CFT$_{p-1}$ at each point in the transverse space of dimension $d-p+1$ \cite{Faulkner:2009wj,DHoker:2009mmn}. In momentum space, this translates into a family of $(p-1)$-dimensional CFT's labeled by the transverse momentum $\vec{k}_{d-p+1}$. Examples with $p=2$ include the charged AdS Reissner-Nordstr\"om (RN) black brane, with $\m$ the chemical potential, and the neutral solution of \cite{Bardoux:2012aw,Andrade:2013gsa}, with $\m$ the source for the translation-symmetry breaking axions. The magnetized CFT$_2$ of \cite{DHoker:2009mmn} is an example with $p=3$, where $\m$ is related to the magnetic field via $\m = \sqrt{B}$. 

The black brane metrics may be written in conformal coordinates as
\begin{equation}
\label{neh0} \mathrm{ds}^2 = \ex^{2A(r)}\left( -f(r)\intd t^2 + f(r)^{-1}\intd r^2 + \intd s_{d-1}^2 \right) \, ,
\end{equation}
with $A(r)$ the scale factor and $f(r)$ the blackening function. $f(r)$ vanishes at the horizon radius $r_H$, which is a function of $T$ and $\m$. In the near-extremal limit, $r_H$ approaches its extremal value
\begin{equation}
\label{neh0b} r_H(T,\m) = r_e(\m)\left(1 + \OO\left(\frac{T}{\mu}\right)\right) \, ,
\end{equation}
where the extremal radius $r_e$ is proportional to $\m^{-1}$.

We now consider two-sided correlators on these black brane geometries, in a specific low energy regime 
\begin{equation}
\label{neh1} |\omega|,|\vec{k}|,T \ll r_e^{-1} \, ,   
\end{equation}
that we call \textit{near-extremal hydrodynamics}. In this regime, the poles of the correlators may be divided into three categories (see figure \ref{Fig:NEpoles} for an illustration):
\begin{itemize}
\item Hard poles $\Omega_n(\vec{k}) = \OO(r_e^{-1})$;
\item Soft poles $\omega^s_n(\vec{k}) = \OO(T)$;
\item Gapless poles $\omega^g_n(\vec{k})$, which go to zero at zero momentum. Such poles were exhibited in \cite{Edalati:2010hk,Edalati:2010pn,Davison:2013bxa,Moitra:2020dal}. They do not go to zero in the limit of vanishing temperature\footnote{With this definition, the results of \cite{Davison:2022vqh} indicate that the gapless poles (and the soft poles) are generically discontinuous as a function of momentum near a collision with a soft pole. The discontinuities are small, of order $\OO(r_e T)$.}, where they instead follow a dispersion relation which at leading order in momentum is similar to hydrodynamic modes\footnote{Because of this similarity, and following the terminology of \cite{Blake:2017ris}, a possible name for these modes would be ``quantum hydrodynamic modes". They clearly differ from standard hydrodynamic modes since they contain non-analytic logarithmic terms at higher order in $r_e|\vec{k}|$ \cite{Gouteraux:2025kta}.} (e.g. a sound mode $\o^g(\vec{k})\sim c_s|\vec{k}|-i\G \vec{k}^2+\OO(r_e^3\vec{k}^4)$ \cite{Edalati:2010pn,Moitra:2020dal} or diffusive mode $\o^g(\vec{k})\sim-iD\vec{k}^2+\OO(r_e^3\vec{k}^4)$ \cite{Edalati:2010hk,Davison:2013bxa,Moitra:2020dal}).
\end{itemize}

Now, if we were in the standard hydrodynamic regime
\begin{equation}
\label{neh2} |\omega|,|\vec{k}|\ll T \ll r_e^{-1} \, ,
\end{equation}
the correlators would be dominated by the gapless (hydrodynamic) poles, whereas the product over the hard and soft poles in \eqref{pf1} would be effectively analytic\footnote{The inverse products over the hard and soft poles are meromorphic functions, which means that they are analytic away from their poles. Since the soft poles are of order $\OO(T)$ and the hard poles of order $\OO(r_e^{-1})$, this implies in particular that these products are analytic in the region of the standard hydrodynamic regime $\omega,k\ll T\ll r_e^{-1}$.}, and resum into the residues of the hydrodynamic poles 
\begin{align}
\nn G_{12}(\omega,\vec{k}) &= \frac{\gamma(\vec{k})}{\prod_{n\in\text{gapless}}\left(1-\frac{\omega^2}{\omega_n^g(\vec{k})^2}\right)\left(1-\frac{\omega^2}{(\omega_n^{g*}(\vec{k}))^2}\right)}\times \\
\nn &\hp{=}\times \frac{1}{\prod_{n=1}^\infty\left(1-\frac{\omega^2}{\omega_n^s(\vec{k})^2}\right)\left(1-\frac{\omega^2}{(\omega_n^{s*}(\vec{k}))^2}\right)
\prod_{n=1}^\infty\left(1-\frac{\omega^2}{\Omega_n(\vec{k})^2}\right)\left(1-\frac{\omega^2}{(\Omega_n^*(\vec{k}))^2}\right)}\\
\label{neh3} &= \frac{\tilde{\gamma}(\omega,\vec{k})}{\prod_{n\in\text{gapless}}\left(\omega^2-\omega_n^g(\vec{k})^2\right)\left(\omega^2-(\omega_n^{g*}(\vec{k}))^2\right)} \, ,
\end{align}
with $\tilde{\gamma}$ analytic for $|\omega|,|\vec{k}|\ll T$. 

By contrast, in the general near-extremal limit \eqref{neh1}, both gapless and soft poles contribute to the effective correlator  
\begin{align}
\label{neh4} G_{12}(\omega,\vec{k}) = \frac{\tilde{\gamma}_e(\omega,\vec{k})\GG^s_{12}(\o,\vec{k})}{\prod_{n\in\text{gapless}}\left(\omega^2-\omega_n^g(\vec{k})^2\right)\left(\omega^2-(\omega_n^{g*}(\vec{k}))^2\right)} \, ,
\end{align}
where $\tilde{\gamma}_e$ is analytic for $|\omega|,|\vec{k}|\ll r_e^{-1}$, and we defined the soft factor as
\begin{equation}
\label{neh4b} \GG^s_{12}(\o,\vec{k}) \equiv \frac{1}{\prod_{n=1}^\infty\left(1-\frac{\omega^2}{\omega_n^s(\vec{k})^2}\right)\left(1-\frac{\omega^2}{(\omega_n^{s*}(\vec{k}))^2}\right)} \, .
\end{equation}

At this point it is interesting to consider the limit $T\to 0$, for which the structure of the soft sector \eqref{neh4b} is expected to simplify. For $|\o|,|\vec{k}|\ll r_e^{-1}$ and away from gapless poles, holographic near-extremal correlators computed from \eqref{pf2b} can indeed be shown to be proportional to the corresponding IR correlators \cite{Faulkner:2009wj,DHoker:2010xwl}\footnote{We generalize the results of \cite{Faulkner:2009wj,DHoker:2010xwl} to general near-extremal black holes in appendix \ref{sec::AppA}. Here we are considering the regime where not only $r_e |\o|$ but also $r_e|\vec{k}|$ is small.}
\begin{equation}
\label{neh4c} G_{12}(\o,\vec{k}) = a(\vec{k}) \GG_{12}(\o,\vec{k}) \left(1+\OO(r_e\o,r_e^2\vec{k}^2)+ \text{gapless poles corrections} \right)  \, ,
\end{equation}
with $\GG_{12}$ the two-sided correlator in the IR CFT$_{p-1}$, and $a(\vec{k})$ some function of the momentum. At zero temperature, the IR correlator is constrained by conformal symmetry to take the form
\begin{equation}
\label{neh4d} \GG_{12}(\o,\vec{k}) = \frac{\mathcal{C}(\vec{k}_{d-p+1})}{\sinh(\b\o/2)} \big(r_e^2(\o^2-\vec{k}_{p-2}^2)\big)^{\D(\vec{k}_{d-p+1})-(p-1)/2} \, ,
\end{equation}
where the $\sinh$ comes from \eqref{pf3}, and we split the spatial momentum into the part that is along the CFT direction $\vec{k}_{p-2}$ and the transverse part $\vec{k}_{d-p+1}$. $\D(\vec{k}_{d-p+1})$ is the IR conformal dimension associated to $G_{12}$. For generic values of the momentum, $\D$ is not an integer, which implies that the correlator $G_{12}$ has branch points at $\omega = \pm |\vec{k}_{p-2}|$ (there is a single branch point at $\omega=0$ for $p=2$). This means that the soft poles merge into branch cuts in the zero-temperature limit \cite{Faulkner:2009wj}. 

From the above observations, we deduce that for $T\to 0$, the soft factor \eqref{neh4b} should become free of poles, with branch points instead at $\omega = \pm |\vec{k}_{p-2}|$, near which the numerator of \eqref{neh4} behaves as 
\begin{equation}
\label{neh4e} \tilde{\g}_e(\o,\vec{k}) \GG^s_{12}(\o,\vec{k}) = \frac{\tilde{a}(\vec{k})}{\sinh(\b\o/2)} \big(r_e^2(\o^2-\vec{k}_{p-2}^2)\big)^{\D(\vec{k}_{d-p+1})-(p-1)/2} \left(1+\OO(r_e\o,r_e^2\vec{k}^2)\right) \, .
\end{equation}
Note that, unlike \eqref{neh4c}, the series in \eqref{neh4e} is expected to remain well defined for any $(\o,\vec{k})$ such that $r_e|\o|,r_e|\vec{k}|\ll 1$. In particular, the expansion does not break down when $\o$ approaches gapless poles, since the latter have been factored out in \eqref{neh4}. However, due to the presence of the branch points, typical terms in the series will contain logarithms, where both $\log{(r_e^2(\o^2-\vec{k}_{p-2}^2))}$ and $\log{(r_e^2\vec{k}_{d-p+1}^2)}$ may in principle appear (appendix \ref{sec::AppB} gives an example with $p=2$). Finally, rewriting \eqref{neh4e} in terms of a generalized susceptibility $\g_e(\o,\vec{k})$, and using \eqref{pf3}, for $|\o|,|\vec{k}|\gg T$ the product formula \eqref{neh4} may be written as 
\begin{equation}
\label{neh4x} \text{Im}G_R(\omega,\vec{k}) = \frac{\g_e(\omega,\vec{k}) \big(r_e^2(\o^2-\vec{k}_{p-2}^2)\big)^{\D(\vec{k}_{d-p+1})-(p-1)/2}}{\prod_{n\in\text{gapless}}\left(\omega^2-\omega_n^g(\vec{k})^2\right)\left(\omega^2-(\omega_n^{g*}(\vec{k}))^2\right)} \, .
\end{equation}
In the following, we refer to \eqref{neh4x} as the \textit{extremal hydrodynamic product formula}. 

The product formula \eqref{neh4x} thus exhibits an interesting structure for the two-sided correlator, where both the gapless poles and the non-analytic IR behavior appear explicitly, moreover in a simple, factorized form. The gapless poles and the generalized susceptibility $\g_e$ are expected to admit expansions as generalized hydrodynamic series, including both integer powers and logarithms of energy and momentum. 

In particular, it is an important aspect of \eqref{neh4x} that it allows to extend the near-extremal hydrodynamic expansion to the region near the branch points. In a standard expansion, the power-law \eqref{neh4e} would indeed be expanded at low momentum, generating terms of the form\footnote{An example is given by the result of \cite{Gouteraux:2025kta} discussed in appendix \ref{sec::AppB}, which contains the first term in this expansion.} $\vec{k}_{d-p+1}^{2m}\log{(\o^2-\vec{k}_{p-2}^2)}$, which imply a breakdown of the expansion at the branch points. The leading behavior of the numerator \eqref{neh4e} resums all such terms, which allows to include the branch points in the near-extremal hydrodynamic regime. Section \ref{sec:application} gives a concrete example of this IR improvement in the case of $p=2$.     

Another comment is related to the number of gapless poles that appear in the denominator of \eqref{neh4}, which may actually depend on the order of the hydrodynamic expansion. Appendix \ref{sec::AppB} gives an example of this, where the single leading order diffusive mode $\o_D(\vec{k})=-iD\vec{k}^2$, splits into two poles with finite real parts at next-to-leading order \cite{Gouteraux:2025kta}. In this case $\gamma_e(\omega,\vec{k})$ should have zeroes at the appropriate order to account for the pole merging.   

\subsection{Special behavior of the soft poles}

\label{sec:special}

In some cases, numerical results show that the soft poles approach precisely the IR poles in the extremal limit $r_e T\to 0$
\begin{equation}
\label{neh4f} \omega^s_n(\vec{k}) = \omega^{\text{IR}}_n(\vec{k})\left(1+ \OO(r_e T)\right) \, ,
\end{equation}
where $\omega^{\text{IR}}_n$ are the poles of the IR CFT$_{p-1}$ correlator $\GG_{12}$. Examples of this behavior for $p=2$ are given in \cite{Edalati:2010hk,Berkooz:2012qh,Davison:2022vqh,Jia:2024zes,Arnaudo:2024sen} (with \cite{Gouteraux:2025kta} being a counter-example), whereas section \ref{sec:mb} describes an example with $p=3$. Based on known examples, it looks like \eqref{neh4f} typically holds either for correlators that do not have gapless poles (e.g. the transverse current in section \ref{sec:ccc} or the scalar in section \ref{sec:mb}), or that couple to energy and momentum (e.g. the correlators of section \ref{sec:sound}).  

When \eqref{neh4f} holds, for all $\omega$ far enough from the IR poles, the soft factor \eqref{neh4b} can be directly expressed in terms of $\GG_{12}$. This can be seen from the fact that $\GG_{12}$ itself is a holographic thermal correlator for which a product formula \eqref{pf1} may be written
\begin{equation}
\label{neh5} \GG_{12}(\omega,\vec{k}) = \frac{\mathcal{K}(\vec{k})}{\prod_{n=1}^\infty\left(1-\frac{\omega^2}{\omega_n^{\text{IR}}(\vec{k})^2}\right)\left(1-\frac{\omega^2}{(\omega_n^{\text{IR}*}(\vec{k}))^2}\right)} \, ,
\end{equation}
with $\mathcal{K}(\vec{k})$ non-zero. Note that here it is important that the IR correlator is actually free of zeroes\footnote{In particular, since the IR correlator can be computed from the $z\to\infty$ limit of the fluctuation equation \eqref{pf2b} (see appendix \ref{sec::AppA}), it is clear that if $G_{12}$ obeys the no-zeroes conditions, then so does $\GG_{12}$. It is also possible that the IR correlator $\GG_{12}$ is free of zeroes even though $G_{12}$ is not.}. Then, if $\omega$ is such that for all IR modes $|\omega-\omega_n^{\text{IR}}|/T\gg r_eT$ (including in particular physical energies $\omega\in\mathbb{R}$), then the product over the soft poles can be approximated in the near-extremal limit by
\begin{equation}
\label{neh5b}  \frac{1}{\prod_{n=1}^\infty\left(1-\frac{\omega^2}{\omega_n^s(\vec{k})^2}\right)\left(1-\frac{\omega^2}{(\omega_n^{s*}(\vec{k}))^2}\right)} = \mathcal{K}(\vec{k})^{-1} \GG_{12}(\omega,\vec{k}) (1+\OO(r_e T)) \, .   
\end{equation}
Substituting into \eqref{neh4}, and defining $\tilde{\g}_e(\o,\vec{k})\equiv \g_e(\o,\vec{k})/\mathcal{K}(\vec{k})$, makes the low-energy correlator directly appear in the expression of $G_{12}$
\begin{equation}
\label{neh6}  G_{12}(\omega,\vec{k}) = \frac{\tilde{\g}_e(\omega,\vec{k})}{\prod_{n\in\text{gapless}}\left(\omega^2-\omega_n^g(\vec{k})^2\right)\left(\omega^2-(\omega_n^{g*}(\vec{k}))^2\right)}\GG_{12}(\omega,\vec{k})(1+\OO(r_e T)) \, , 
\end{equation}
which is valid for all $\omega$ outside small disks of radius of order $\OO(r_e T)$ around the IR poles. The generalized susceptibility $\tilde{\gamma}_e(\omega,\vec{k})$ is analytic in the near-extremal hydrodynamic regime \eqref{neh1}.

In the general near-extremal regime \eqref{neh1}, $\GG_{12}(\o,\vec{k})$ is the two sided correlator in the IR CFT$_{p-1}$ at finite temperature. This makes equation \eqref{neh6} particularly useful in the case of $p = 2$ and 3, since the general expression of thermal two-point functions is known for one-dimensional and two-dimensional CFT's (see e.g. \cite{Faulkner:2009wj} and \cite{Dodelson:2023vrw,Son:2002sd}). In these cases, the soft factor \eqref{neh5b} can therefore be expressed at leading order in $r_e T$ in a closed form (up to a $\vec{k}$-dependent factor), including for energies $|\o|\lesssim T$.  

This concludes the general presentation of the near-extremal hydrodynamic product formula. The next section discusses some concrete examples.

\section{Examples of product formulae}
\label{sec:Ex}

We present in this section several examples where the product formula applies and provides useful expressions for different types of holographic correlators. This includes cases with different gapless modes -- both diffusive (section \ref{sec:ccc}) and sound modes (section \ref{sec:sound}) -- and for different values of $p$: $p=2$ (sections \ref{sec:ccc} and \ref{sec:sound}) and $p=3$ (section \ref{sec:mb}). 

\subsection{Probe current correlators}
\label{sec:ccc}

We consider here a first example, corresponding to gauge field fluctuations $A_\m$, which obey the Maxwell equations
\begin{equation}
\label{E11} \partial_M\left(\sqrt{-g}F^{MN}\right) = 0 , \qquad F_{MN} \equiv \pa_M A_N - \pa_N A_M\, ,
\end{equation} 
on top of the Reissner-Nordstr\"om background\footnote{All results in this section are actually independent of the blackening function $f(r)$, as long as it admits an AdS$_2$ IR geometry in the near-extremal limit.} with metric 
\begin{equation}
\label{E11b} \intd s^2 = \frac{\ell^2}{r^2}\left(-f(r)\intd t^2 + f(r)^{-1}\intd r^2 + \intd \vec{x}^2\right) \, , 
\end{equation}
where the blackening function $f(r)$ vanishes at the horizon radius $r_H$.

The operator dual to $A_\m$ is a conserved current $J^\mu$, with vanishing background density $\left<J^0\right>=0$, and not charged under any other background density. This corresponds to the correlators studied in \cite{Jarvinen:2023xrx}, that are related to neutrino transport.  
 
At finite momentum $\vec{k}$, the perturbations split into transverse and longitudinal modes, such that the Fourier space current-current correlator may be written as
\begin{equation}
\label{E12} \left<J_\m J_\n\right>(\omega,\vec{k}) = P_{\m\n}^\perp(\omega,\vec{k}) i\Pi^\perp(\omega,\vec{k}) + P_{\m\n}^\parallel(\omega,\vec{k}) i\Pi^\parallel(\omega,\vec{k}) \, , 
\end{equation}
with the non-zero components of the transverse and longitudinal projectors respectively given by 
\begin{equation}
\label{E13} P_{ij}^\perp(\omega,\vec{k}) = \d_{ij}-\frac{k_ik_j}{\vec{k}^2}\, .
\end{equation}
\begin{equation}
\label{E14} P_{tt}^\parallel(\omega,\vec{k}) = \frac{\vec{k}^2}{\omega^2 - \vec{k}^2} , \quad P_{ti}^\parallel(\omega,\vec{k}) = -\frac{\omega k^i}{\omega^2-\vec{k}^2} , \quad P_{ij}^\parallel(\omega,\vec{k}) = \frac{\omega^2}{\omega^2-\vec{k}^2}\frac{k_ik_j}{\vec{k}^2} \, .
\end{equation}
The polarization functions $\Pi^\perp$ and $\Pi^\parallel$ may be computed by solving the decoupled Maxwell equations for the two types of modes 
\begin{equation}
\label{E15} \pa_r\left(\frac{f(r)}{r^{d-3}}\pa_r A_i^\perp\right) + \frac{1}{r^{d-3}f(r)}\left(\omega^2 - f(r)\vec{k}^2\right)A_i^\perp = 0 \, ,
\end{equation}
\begin{equation}
\label{E16} \pa_r\left(\frac{f(r)}{r^{d-3}(\omega^2-f(r)\vec{k}^2)}\pa_r E^\parallel\right) + \frac{1}{r^{d-3}f(r)}E^\parallel = 0 \, ,
\end{equation}
where $d\geq 3$ is the boundary dimension\footnote{We focus on theories that feature diffusion, which excludes in particular (1+1)-dimensional CFT's \cite{Kovtun:2008kx,Herzog:2007ij}.} and the transverse gauge field and longitudinal electric field are defined as
\begin{equation}
\label{E17} A_i^\perp \equiv A_i - \hat{k}_i\hat{k}^jA_j ,\quad E^\parallel \equiv |\vec{k}| A_t + \omega \hat{k}^jA_j, \qquad \hat{k}^i = \frac{k^i}{|\vec{k}|} \, .
\end{equation}

Equations \eqref{E15} and \eqref{E16} are not in the Schr\"odinger form \eqref{pf2b}. \eqref{E15} can be put in the appropriate form by going to tortoise coordinates and with a simple field redefinition
\begin{equation}
\label{E18} A_i^\perp(r) \to r(z)^{(d-3)/2} \psi^\perp(z), \quad \frac{\intd z}{\intd r} = \frac{1}{f(r)} \, ,
\end{equation}
which gives 
\begin{equation}
\label{E19} \pa_z^2\psi^\perp + (\omega^2 - V^\perp(z))\psi^\perp(z) = 0 \, ,
\end{equation}
\begin{equation}
\label{E110} V^\perp(z) = f(z)\vec{k}^2 + \frac{(d-3)(d-1)f(z)^2}{4r(z)^2} - \frac{(d-3)f'(z)}{2r(z)} \, .
\end{equation}
The potential \eqref{E110} is regular for $z\in(0,\infty)$, and its UV behavior is as in \eqref{pf2d} with $\n^2 = (d-2)^2/4 \geq 0$. The transverse two-sided polarization function is therefore free of zeroes.  

The longitudinal equation \eqref{E16} can be put into the Schr\"odinger form with a different field redefinition
\begin{equation}
\label{E111} E^\parallel(r) \to \left(r(z)^{d-3}(\omega^2-f(z)\vec{k}^2)\right)^{1/2} \psi^\parallel(z) \, ,
\end{equation}
with the corresponding potential given by 
\begin{align}
\nn V^\parallel(z) = &f(z)\vec{k}^2 + \frac{(d-3)(d-1)f(z)^2}{4r(z)^2} - \frac{(d-3)f'(z)}{2r(z)} 
-\frac{(d-3)\vec{k}^2 f(z) f'(z)}{2 r(z) \left(\omega ^2- f(z)\vec{k}^2\right)}+\\
\label{E112} &+\frac{3 \vec{k}^4 f'(z)^2}{4 \left(\omega ^2-f(z)\vec{k}^2 \right)^2} +\frac{\vec{k}^2 f''(z)}{2 \left(\omega ^2-f(z)\vec{k}^2 \right)} \, .   
\end{align}
This potential has a pole when $f(z)\vec{k}^2=\omega^2$, which can be reached at finite $z$ for $\o^2/\vec{k}^2 < 1$. Also its UV asymptotics are such that $\n^2 = (d-2)^2/4$, except when $\o^2 = \vec{k}^2$, for which $\n^2 = (d-1)^2$. As a consequence, we expect the corresponding two-sided correlator to feature some zeroes for $\o^2/\vec{k}^2\leq 1$. It can be checked (see e.g. \cite{Jarvinen:2023xrx}) that the correlator associated with $E^\parallel$ is the polarization function $\Pi^\parallel$ in \eqref{E12}. In this case, it is therefore simple to understand why there is a zero: since the full correlator should not have a pole at $\omega^2=\vec{k}^2$, and because $\Pi^\parallel$ is multiplied by the longitudinal projector \eqref{E13}, it results that $\Pi^\parallel$ should vanish when $\omega^2=\vec{k}^2$. 

We will now check that there is no other zero than $\omega^2 = \vec{k}^2$, which can be done by considering the field
\begin{equation}
\label{E113} \Psi(z) \equiv \frac{\pa_zE^\parallel}{r(z)^{(d-3)/2}(\omega^2-f(z)\vec{k}^2)} \, .
\end{equation}
The latter obeys 
\begin{equation}
\label{E114} \Psi''(z) + (\omega^2-V_\Psi(z))\Psi(z) = 0\, ,
\end{equation}
\begin{equation}
\label{E115} V_\Psi(z) = f(z)\vec{k}^2 + \frac{(d-3)(d-5)f(z)^2}{4r(z)^2} + \frac{(d-3)f'(z)}{2r(z)} \, ,
\end{equation}
so the corresponding two-sided correlator $G_\Psi$ is free of zeroes. Since $G_\Psi$ is related to the two-sided polarization function as\footnote{See appendix \ref{sec::AppB2} for details.} 
\begin{equation}
\label{E116} \Pi_{12}^\parallel(\omega,\vec{k}) = c_0(\omega^2-\vec{k}^2) G_\Psi(\omega,\vec{k}) \, ,
\end{equation}
with $c_0$ a constant, we find that the only zero of the polarization function is indeed at $\omega^2 = \vec{k}^2$. 

\subsubsection{(Near-)extremal hydrodynamic limit}

\label{sec:ccneh}

In the limit $T\ll|\o|,|\vec{k}|\ll r_e^{-1}$ (with $r_e$ the extremal horizon radius), the extremal hydrodynamic product formula \eqref{neh4x} applies to $G_\Psi$, with $p=2$ and the IR scaling dimension given by
\begin{equation}
\label{E118} \D(\vec{k}) \equiv \frac{1}{2}+ \frac{1}{2}\sqrt{1+\frac{8\vec{k}^2}{\pa^2_r f(r_H)}} \, .
\end{equation}
At leading order in the near-extremal hydrodynamic expansion, there is a single gapless pole $\omega_D(\vec{k}) = -iD\vec{k}^2$, with $D=r_H/(d-2)$ the diffusion constant \cite{Davison:2013bxa}. In this regime, the longitudinal spectral function may thus be written as 
\begin{equation}
\label{E120} \text{Im}\Pi_R^\parallel(\omega,\vec{k}) = -(\omega^2-\vec{k}^2)\frac{\s\o^{2\D(\vec{k})-1}}{\omega^2+D^2\vec{k}^4} \, ,
\end{equation}
with $\s$ the conductivity. The expression \eqref{E120} corresponds to the standard hydrodynamic expression for a diffusive mode \cite{Jarvinen:2023xrx}, improved with the appropriate IR power-law. 

At next-to-leading order, the numerical analysis presented in appendix \ref{sec::AppB3} indicates that $\o_D$ splits into two hydrodynamic poles, much like in \cite{Gouteraux:2025kta} (see \eqref{B6}). At this order and higher, we therefore expect the polarization function to be described by an expression of the form \eqref{B7} in the extremal hydrodynamic regime. 

We now come back to the transverse correlator, which does not admit any zeroes nor gapless poles. In this case, the near-extremal product formula \eqref{neh4} therefore predicts that $\Pi^\perp$ should factorize as
\begin{equation}
\label{E122} \Pi^\perp_{12}(\omega,\vec{k}) = \g(\omega,\vec{k}) \GG^s_{12}(\omega,\vec{k}) \, ,
\end{equation}
with $\g$ analytic and free of zeroes, and $\GG^s_{12}$ the soft factor \eqref{neh4b}. A numerical analysis of the transverse fluctuation equation \eqref{E19} further indicates that the scenario discussed in section \ref{sec:special} applies in this sector\footnote{In the longitudinal sector figure \ref{Fig:Impoles_Pil} indicates that, although the soft poles are also parametrically close to the IR poles away from collisions with the gapless pole, they receive corrections of order 1 near such collisions, as was also observed in \cite{Gouteraux:2025kta}. This is why the general near-extremal product formula \eqref{neh6} does not apply in all generality for the longitudinal correlator. That being said, figure \ref{Fig:Impoles_Pil} also indicates that \eqref{neh6} applies to the longitudinal correlator in the regime $(r_e \vec{k})^2\ll r_eT$. It should also be a good approximation for $(r_e \vec{k})^2\gtrsim r_eT$ as long as $|\omega| \ll r_e\vec{k}^2$. In the latter case, the soft poles that receive order 1 corrections from collisions with the gapless pole -- which are of the same order as the gapless pole according to figure \ref{Fig:Impoles_Pil} -- indeed give a small contribution to the full product over the soft poles \eqref{neh4b} (and hence to the correlator \eqref{neh4}).}, with the soft poles approaching the IR poles in the limit $r_e T\to 0$ (see appendix \ref{sec::AppB3}). The near-extremal transverse polarization function may thus be written in the form \eqref{neh6}, i.e. an analytic factor times the IR AdS$_2$ correlator \cite{Faulkner:2009wj}
\begin{equation}
\label{E117} \GG_{12}(\omega,\vec{k}) = (\pi T)^{2\D-1}\frac{\cos{(\pi(\D-1))}}{\pi}\frac{\Gamma\left(\frac{3}{2}-\D\right)}{\Gamma\left(\frac{1}{2}+\D\right)}\left|\G\left(\D-\frac{i\o}{2\pi T}\right)\right|^2 \, .
\end{equation}
Absorbing for convenience some of the $\vec{k}$-dependent terms in \eqref{E117} into the analytic factor, we get 
\begin{equation}
\label{E122b} \Pi^\perp_{12}(\omega,\vec{k}) = -\s^\perp(\omega,\vec{k}) \frac{(2\pi T)^{2\D-1}}{2\pi}\left|\G\left(\D-\frac{i\o}{2\pi T}\right)\right|^2(1+\OO(r_e T)) \, ,
\end{equation}
with $\s^\perp$ analytic in the near-extremal hydrodynamic regime.

In the extremal limit $|\omega|\gg T$, we recover the extremal hydrodynamic product formula  
\begin{equation}
\label{E121} \text{Im}\Pi^\perp_R(\omega,\vec{k}) = -\s^\perp(\omega,\vec{k}) \o^{2\D(\vec{k})-1} \, .
\end{equation}
In particular, at leading order in the hydrodynamic expansion we get
\begin{equation}
\label{E123} \text{Im}\Pi^\perp_R(\omega,\vec{k}) = -\s \o^{2\D(\vec{k})-1}(1+\OO(r_e\o,r_e^2\vec{k}^2)) \, ,
\end{equation}
where we used that the two polarization functions should be equal for $\vec{k}=0$ to get $\s^\perp(0)=\s$, with $\s$ the same conductivity that appears in the expression of the longitudinal correlator \eqref{E120}. 

The leading order extremal hydrodynamic expressions \eqref{E120} and \eqref{E123} will be compared with the numerically computed exact correlators in section \ref{sec:application}. Before that, the following subsections discuss other settings where the product formula gives interesting information about the structure of the correlator in the near-extremal hydrodynamic regime.

\subsection{Stress-tensor correlators in the sound channel}

\label{sec:sound}

We now consider the case of helicity-0 stress-tensor correlators on the Reissner-Nordtr\"om (RN) background, whose gapless sector is qualitatively different from the previous example, since it features two types of gapless modes: a diffusive-like mode $\omega_D(\vec{k}) = -iD\vec{k}^2+\OO(\vec{k}^4)$, and a sound-like mode $\omega_{s}(\vec{k}) = c_s |\vec{k}| - i\Gamma \vec{k}^2 + \OO(\vec{k}^4)$ \cite{Edalati:2010pn,Davison:2011uk,Moitra:2020dal}.  

The background metric is given  by
\begin{equation}
\label{E20a} \intd s^2 = \frac{\ell^2}{r^2}\left(-f(r)\intd t^2+f(r)^{-1}\intd r^2 + \intd\vec{x}^2\right) \, ,
\end{equation}
\begin{equation}
\label{E20b} f(r) = 1 -2Mr^{n+1} + Q^2r^{2n} \, ,
\end{equation}
where $n\equiv d-1\geq 2$ is the number of spatial dimensions at the boundary, that are denoted $x^i=(x,y_1,\dots,y_{n-1})$. The RN solution also contains a background gauge field 
\begin{equation}
\label{E20} A = \frac{q}{n-1}\left(r_H^{n-1} - r^{n-1}\right)\intd t \, ,
\end{equation}
with $q$ proportional to the density for the current dual to $A$, and $r_H$ the horizon radius.

The stress-tensor correlators are computed from the fluctuation equations for the appropriate modes of the metric $g_{MN}$, which couple to the fluctuations of the gauge field \eqref{E20}. These equations can be written in terms of two gauge-invariants \cite{Kodama:2003kk,Edalati:2010pn}
\begin{equation}
\label{E21} \mathcal{A}(r) \equiv r^{2-n} \pa_r \d A_t+\frac{q}{2}\left(\d g^t_t-\frac{n}{n-1}(\d g ^i_i-\d g^x_x)\right) \, ,
\end{equation}
\begin{equation}
\label{E21b} \Phi(r) \equiv r^{-n/2}\left[\d g^i_i-\d g^x_x+ \frac{(n-1)f(r)}{H(r)}r\pa_r\d g^i_i\right] \, ,
\end{equation}
where $\d\varphi$ refers to a small fluctuation of the field $\varphi$, and we took the 3-momentum $\vec{k}$ to lie in the $x$ direction (without loss of generality); we also defined 
\begin{equation}
\label{E21c} H(r) \equiv r^2\vec{k}^2 - \frac{n}{2}rf'(r) \, .
\end{equation}
The fields $\AA$ and $\Phi$ obey \cite{Kodama:2003kk,Edalati:2010pn}
\begin{align}
\label{E22} 
& f(r)r^{2-n}\pa_{r}\left( \frac{f(r)}{r^{2-n}}\pa_{r}\AA \right)
   +\left(\omega^2-f(r)\vec{k}^2
        -\frac{2n^2(n-1)\zeta(r)f(r)^2}{r^2H(r)}\right)\AA 
\nn\\
& \qquad\qquad    
   =n(n-1)f(r)r^{n/2-2}q\left( \frac{4H(r)^2-nP_Z(r)}{4nH(r)}\Phi
      -f(r)r\partial_{r}\Phi \right)  \, ,
\end{align}
\begin{equation}
\label{E22b} 
f(r)\pa_r\left(f(r)\pa_r\Phi\right) +(\omega^2-V_S(r))\Phi = (n-1)\frac{Q^2}{q} \frac{f(r)P_S(r)}{H(r)^2}r^{3n/2-2}\AA \, ,
\end{equation}
where the expressions for the coefficients $\zeta(r),P_Z(r),P_S(r)$ and $V_S(r)$ are listed in appendix \ref{sec::AppC}. 

It is possible to further simplify the problem by introducing the following ``master" variables\footnote{Note the relations between our variables and those of Kodama-Ishibashi (KI) \cite{Kodama:2003kk}
\begin{equation}
\nn \AA = \vec{k}^2\AA_{\text{KI}} \quad , \quad \Phi = \frac{1}{n}\vec{k}^2\Phi_{\text{KI}} \, ,
\end{equation}
\begin{equation}
\nn \Phi_+ = \frac{\vec{k}^2}{4nQ} \Phi_{-,\text{KI}} \quad , \quad \Phi_- = -\frac{\vec{k}^2}{(n+1)(M+\r)} \Phi_{+,\text{KI}} \, .
\end{equation}}
\begin{equation}
\label{E23} \Phi_\pm \equiv \a_\pm(r)\Phi -\frac{Q}{q}r^{n/2-1} \AA \, ,
\end{equation}
where 
\begin{equation}
\label{E23b} \a_\pm(r)\equiv (n+1)\frac{M\pm\r(\vec{k})}{4Q}-\frac{n}{2}Qr^{n-1} \quad , \quad \r(\vec{k}) \equiv \sqrt{M^2+\frac{4\vec{k}^2Q^2}{(n+1)^2}} .
\end{equation}
These variables obey decoupled equations
\begin{equation}
\label{E24} f(r)\pa_r\left(f(r)\pa_r\Phi_\pm\right) +(\omega^2-V_\pm(r))\Phi_\pm = 0  \, ,
\end{equation}
where the potential $V_\pm$ can be expressed as
\begin{align}
\nn V_\pm(r) =& \pm \frac{2Q}{(n+1)\rho(\vec{k})}\bigg\{\a_\pm(r)V_S(r)+\\
\nn &+\frac{f(r)}{4nr^3H(r)}\bigg[2 n(n-1) Q r^n \bigg( n^2 r f'(r)\Big(rf'(r)- 2(2n+1)f(r)\Big)-\\
\nn &-4n(n-2)f(r)\vec{k}^2r^2 + 4 n^3 (n+1) f(r)(f(r)-1) -4\vec{k}^4 r^4\bigg)+\\
\nn&+\frac{n}{2} r \alpha_\mp(r) \bigg(r f'(r) \Big(n (5 n (2-3 n)+8) f(r)+ 8 \vec{k}^2 r^2+ 2 n(n-2) r f'(r)\Big) -\\
\label{E25} &-2 f(r) \Big((n-4)(n-2)\vec{k}^2r^2 -8n^2(n^2-1)(f(r)-1)\Big)-8 \vec{k}^4r^4\bigg)\bigg]\bigg\} \, .
\end{align}

Equation \eqref{E24} finally takes the Schr\"odinger form \eqref{pf2b} if we use the tortoise coordinates $z$
\begin{equation}
\label{E26} \pa^2_z\Phi_\pm + (\omega^2-V_\pm(z))\Phi_\pm = 0  \, ,
\end{equation}
with $z'(r)=f(r)^{-1}$. The potentials $V_\pm(r)$ in \eqref{E25} can only diverge at finite $r$ if $H(r) = \vec{k}^2r^2-(n/2)rf'(r)$ goes to zero. Since $f'(r)$ is negative, this can only happen for negative $\vec{k}^2$. So $V_\pm(r)$ is regular at finite $r$ for $\vec{k}^2\notin \mathbb{R}_-^*$. Also the UV asymptotics of the potentials follow \eqref{pf2d} with $\n^2 = (n-3)^2/4$ at finite momentum and $\n^2 = (n+1)^2/4$ at $\vec{k}=0$, which are both positive. We conclude that the variables $\Phi_\pm$ in \eqref{E23} are associated with two-sided correlator $G^\pm_{12}(\omega,\vec{k})$ that obey the product formulae without zeroes, except potentially for negative $\vec{k}^2$.  It would be interesting to better understand to what extent the product formula still applies at $\vec{k}^2<0$\footnote{One could speculate that this singularity is related to the pole-skipping phenomenon, which corresponds to the collision of a gapless pole with a zero at negative values of $\vec{k}^2$ \cite{Grozdanov:2017ajz,Blake:2018leo,Davison:2022vqh}. This does not have to be the case however, since the zero responsible for pole-skipping is already captured by the product formula for the spectral function, due to the $\sinh$ factor in \eqref{pf3}, even if $G_{12}$ itself is free of zeroes.}.  

We would now want to relate the stress-tensor and current correlators to the master fields Green's functions $G^\pm(\omega,\vec{k})$. This relation depends on the dimensionality $n$, so we will now consider a fixed value of $n$. The case of $n=2$ was treated in \cite{Edalati:2010pn}, where the sound channel correlators were found to be non-linearly related to $G^\pm$, as reviewed in appendix \ref{sec::AppD}. Here we will instead consider $n=3$, for which the relation becomes linear. We argue in appendix \ref{sec::AppD} that this also holds for any $n\geq 4$. 

The $G^\pm$ correlators appear in the near-boundary expansion of the solutions to \eqref{E26}. The form of this expansion can be inferred from the behavior of $V_\pm(z)$ at $z\to 0$\footnote{$V_+$ has a special behavior for $\vec{k}=0$: $V_+(z)\sim n(n+2)/(4z^2)$. This case requires a special treatment, but can be computed by taking the limit of the finite $\vec{k}$ result.}
\begin{equation}
\label{E27} V_\pm(z) \underset{z\to 0}{\sim} \frac{(n-3)^2-1}{4z^2} \, ,
\end{equation}
which implies 
\begin{align}
\label{E28} & \underline{n \neq 3}: \Phi_\pm(z) = \Phi_\pm^{(0)}z^{\frac{4-n}{2}}(1+\dots) + \Phi_\pm^{(1)}z^{\frac{n-2}{2}}(1+\dots) \, , \\
\label{E29} & \underline{n = 3}: \Phi_\pm(z) = \Phi_\pm^{(0)}z^{\frac{1}{2}}\log{z}\,(1+\dots) + \Phi_\pm^{(1)}z^{\frac{1}{2}}(1+\dots) \, .
\end{align}
where the dots refer to terms going to zero as $z\to 0$. For a solution of \eqref{E26} with infalling boundary conditions at the horizon, we define the retarded correlator 
\begin{equation}
\label{E210} G^\pm_R(\omega,k) = \frac{\Phi_\pm^{(1)}}{\Phi_\pm^{(0)}} \, ,
\end{equation}
and the two-sided correlator (see \eqref{pf3})
\begin{equation}
\label{E211} G^\pm_{12}(\omega,k) = \frac{\text{Im}\,G^\pm_R(\omega,k)}{\sinh(\beta\omega/2)} \, ,
\end{equation}
which obeys the product formula \eqref{pf1}. 

The stress tensor and current correlators that we are interested in, can them be related to the near-boundary expansion of the metric and gauge field in radial gauge ($n=3$) 
\begin{equation}
\label{E212a} \d g_{\m\n} = \frac{\ell^2}{r^2}\left(\d g_{\m\n}^{(0)} + \d g_{\m\n}^{(2)}r^2 + r^4(\d\pi_{\m\n}+\d \psi_\m\log{r})\right) \, ,
\end{equation}
\begin{equation}
\label{E212b} \d A_\m = \d A_\m^{(0)} + r^2(\d\pi_\m+\d B_\m\log{r}) \, .
\end{equation}
Up to contact terms, the retarded correlators are proportional to the ratio of the vev terms over the sources 
\begin{equation}
\label{E213a} \left<J_\m J_\n\right>_R = C_J\frac{\d\pi_\m}{\d A^\n_{(0)}} + \text{contact terms} \, ,
\end{equation}
\begin{equation}
\label{E213b} \left<T_{\m\n} T_{\a\b}\right>_R = C_T\frac{\d\pi_{\m\n}}{\d g^{\a\b}_{(0)}} + \text{contact terms} \, ,
\end{equation}
\begin{equation}
\label{E213c} \left<T_{\m\n} J_\a\right>_R = C_{\text{mix}}\frac{\d\pi_{\m\n}}{\d A^\a_{(0)}} + \text{contact terms} \, ,
\end{equation}
for solutions of the Einstein-Maxwell equations with infalling boundary conditions at the horizon. $C_J$, $C_T$ and $C_{\text{mix}}$ are constants which only depend on the normalization of the bulk action. Since we are interested in the imaginary part of the retarded correlators, and because the contact terms are real, the latter will not be relevant to our discussion. 

From the definitions \eqref{E23}, and using the near-boundary equations of motion\footnote{The latter imply in particular the Ward identities obeyed by the stress-tensor and current correlators \eqref{E213a}-\eqref{E213c}. See \cite{Edalati:2010pn} for more details of this calculation in the case $n=2$.}, the coefficients $\Phi_\pm^{(0)}$ and $\Phi_\pm^{(1)}$ can now be related to the metric and gauge field expansions \eqref{E212a}-\eqref{E212b} as
\begin{equation}
\label{E214a} \Phi_\pm^{(0)} = -\frac{Q}{q}z_A + \frac{M\pm\r(\vec{k})}{6Q}z_g \, , 
\end{equation}
\begin{align}
\nn \Phi_\pm^{(1)} = &-\frac{2Q}{q}\d\pi_t + \frac{8(M\pm\r(\vec{k}))}{Q\o^2}\d\pi^x_x + \frac{24(M\pm\r(\vec{k}))}{\o^2\vec{k}^2}\left(-\frac{Q}{q}z_A+\frac{M}{6Q}z_g\right) + \frac{M\pm\r(\vec{k})}{12Q}z_g-\\
\label{E214b} &- \frac{Q}{2}\d g^{t(0)}_t + \left(\frac{4M(M\pm\r(\vec{k}))}{Q\o^2}+\frac{3Q}{4}\right)\left(\d g^{y_1(0)}_{y_1}+\d g^{y_2(0)}_{y_2}\right) \, , 
\end{align}
where we defined the combinations of sources
\begin{equation}
\label{E215a} z_A \equiv |\vec{k}|\left(|\vec{k}|\d A_t^{(0)}+\o\d A_x^{(0)}\right) \, ,
\end{equation}
\begin{equation}
\label{E215b} z_g \equiv 2\vec{k}^2\d g^{t(0)}_t - 4\o |\vec{k}|\d g^{x(0)}_t - 2\o^2\d g^{x(0)}_x + (\o^2-\vec{k}^2) \left(\d g^{y_1(0)}_{y_1}+\d g^{y_2(0)}_{y_2}\right) \, .
\end{equation}
The vev terms $\d\pi_t$ and $\d\pi_{xx}$ can then be expressed in terms of the $G^\pm$ correlators by inverting the relations \eqref{E210}, which gives
\begin{equation}
\label{E216a} \d\pi_t = \frac{q}{4Q\r(\vec{k})}\left[(M-\r(\vec{k}))G^+_R\Phi_+^{(0)} - (M+\r(\vec{k}))G^-_R\Phi_-^{(0)}\right] + \text{contact terms} \, ,
\end{equation}
\begin{equation}
\label{E216b} \d\pi_{xx} = \frac{\o^2Q}{16\r(\vec{k})}\left[G^+_R\Phi_+^{(0)} - G^-_R\Phi_-^{(0)}\right] + \text{contact terms} \, .
\end{equation}
Combined with the Ward identities, \eqref{E216a}-\eqref{E216b} finally makes it possible to express the retarded correlators \eqref{E213a}-\eqref{E213c} in terms of $G^\pm_R$
\begin{equation}
\label{E217a} \left<J_aJ_b\right>_R = P_{ab}^\parallel(\omega,\vec{k}) G_J(\omega,\vec{k}) + \text{contact terms} \, ,
\end{equation}
\begin{equation}
\label{E217b} \left<T_{ab}T_{cd}\right>_R = P_{ab}^\parallel(\omega,\vec{k})P_{cd}^\parallel(\omega,\vec{k}) G_T(\omega,\vec{k}) + \text{contact terms} \, ,
\end{equation}
\begin{equation}
\label{E217c} \left<T_{ab}T_{ii}\right>_R = -\frac{1}{2}P_{ab}^\parallel(\omega,\vec{k})G_T(\omega,\vec{k}) + \text{contact terms} \, ,
\end{equation}
\begin{equation}
\label{E217d} \left<(T_{y_1y_1}+T_{y_2y_2})T_{y_1y_1}\right>_R = \left<(T_{y_1y_1}+T_{y_2y_2})T_{y_2y_2}\right>_R = \frac{1}{2} G_T(\omega,\vec{k}) + \text{contact terms} \, ,
\end{equation}
\begin{equation}
\label{E217e} \left<J_{a}T_{bc}\right>_R = \frac{\omega^2-\vec{k}^2}{2\vec{k}^2}\left(P_{ab}^\parallel(\omega,\vec{k})P_{tc}^\parallel(\omega,\vec{k}) + P_{ac}^\parallel(\omega,\vec{k})P_{tb}^\parallel(\omega,\vec{k})\right) G_{\text{mix}}(\omega,\vec{k}) + \text{contact terms} ,
\end{equation}
\begin{equation}
\label{E217f} \left<J_{a}T_{ii}\right>_R = -\frac{\o^2-\vec{k}^2}{2\vec{k}^2}P_{ta}^\parallel(\omega,\vec{k}) G_{\text{mix}}(\omega,\vec{k}) + \text{contact terms} \, ,
\end{equation}
where $a,b,c,d$ take values in $(t,x)$, whereas $i,j$ take values in $(y_1,y_2)$ (``$ii$" is not summed over). We introduced the longitudinal projector transverse to the 2-vector $k^a=(\omega,|\vec{k}|)$
\begin{equation}
\label{E218} P_{ab}^\parallel(\omega,\vec{k}) = \eta_{ab} + \frac{k_ak_b}{\omega^2-\vec{k}^2} \, , 
\end{equation}
and defined the following combinations of master correlators
\begin{equation}
\label{E219a} G_J(\omega,\vec{k}) \equiv C_J\frac{\o^2-\vec{k}^2}{4\r(\vec{k})}\left[(M+\r(\vec{k}))G^-_R(\omega,\vec{k}) - (M-\r(\vec{k}))G^+_R(\omega,\vec{k})\right]\, ,
\end{equation}
\begin{equation}
\label{E219b} G_T(\omega,\vec{k}) \equiv C_T\frac{(\o^2-\vec{k}^2)^2}{48\r(\vec{k})}\left[(M-\r(\vec{k}))G^-_R(\omega,\vec{k}) - (M+\r(\vec{k}))G^+_R(\omega,\vec{k})\right] \, ,
\end{equation}
\begin{equation}
\label{E219c} G_{\text{mix}}(\omega,\vec{k}) \equiv C_{\text{mix}}\frac{q\vec{k}^2(\o^2-\vec{k}^2)}{48\r(\vec{k})}\left[G^+_R(\omega,\vec{k}) - G^-_R(\omega,\vec{k})\right] \, .
\end{equation}

Now taking the imaginary parts on both sides of equations \eqref{E217a}-\eqref{E217f} - and remembering that the contact terms are real - the sound channel two-sided correlators are found to be linearly related to the master correlators $G^\pm_{12}$. Since the latter obey the product formula, the stress tensor and current correlators can therefore also be written as a product over their poles. In particular, the $G^-$ correlator has a diffusive pole $\o_D(k) = -iD\vec{k}^2+\OO(\vec{k}^4)$, whereas $G^+$ features a sound pole $\o_s(\vec{k})= c_s|\vec{k}|-i\G\vec{k}^2+\OO(\vec{k}^4)$ \cite{Arean:2020eus}, as can be seen explicitly in the numerical results presented in appendix \ref{Sec::AppE}. These results also indicate that the soft poles approach the IR poles in the near-extremal limit\footnote{Reference \cite{Davison:2022vqh} showed that the diffusive mode of energy transport $\omega_D(k)$ also interacts with the soft poles when they cross, like the diffusive pole for probe currents shown in figure \ref{Fig:Impoles_Pil}. However, unlike what happened in figure \ref{Fig:Impoles_Pil}, the deviations of the soft poles are very small in the case of energy transport. These corrections are actually not even visible at the scale of figure \ref{Fig:poles_Gm}, which indicates that they should be smaller than order $\OO(r_e T)$. The near-extremal product formula \eqref{neh6} therefore applies to the $G^-$ correlator, despite the pole collisions.}, as in \eqref{neh4f}. Following the discussion of section \ref{sec:special}, in the near-extremal hydrodynamic regime we may therefore write the $G^\pm$ correlators as
\begin{equation}
\label{E220a} G^-_{12}(\omega,\vec{k}) = \frac{\g_-(\o,\vec{k})\GG^-_{12}(\o,\vec{k})}{\o^2-\o_D(\vec{k})^2}\left(1+\OO(r_e T)\right)  \, ,
\end{equation}
\begin{equation}
\label{E220b} G^+_{12}(\omega,\vec{k}) = \frac{\g_+(\o,\vec{k})\GG^+_{12}(\o,\vec{k})}{(\o^2-\o_s(\vec{k})^2)(\o^2-\o_s^
*(\vec{k})^2)}\left(1+\OO(r_e T)\right) \, ,
\end{equation}
with $\g_\pm(\o,\vec{k})$ analytic and free of zeroes (except potentially for $\vec{k}^2<0$), and $\GG^\pm_{12}(\o,\vec{k})$ the IR CFT$_1$ correlators \eqref{E117}, with conformal dimensions
\begin{equation}
\label{E221} \D_\pm(\vec{k}) = \frac{1}{2}\left(1+\sqrt{5+\frac{4\vec{k}^2r_e^2}{n(n+1)}\mp 4\sqrt{1+\frac{4(n-1)\vec{k}^2r_e^2}{n^2(n+1)}}}\right) \, .    
\end{equation}

Then, from \eqref{E219a}-\eqref{E219c} (and using \eqref{pf3}), the imaginary parts of the sound channel retarded correlators may themselves be written as product formulae
\begin{equation}
\label{E222} \text{Im}G_{\text{L}}(\omega,\vec{k}) = \frac{\g_L(\o,\vec{k})\text{Im}\GG^+_R(\o,\vec{k})}{(\o^2-\o_D(\vec{k})^2)(\o^2-\o_s(\vec{k})^2)(\o^2-\o_s^
*(\vec{k})^2)}\left(1+\OO(r_e T)\right) \, , \quad \text{L}\in\{T,J,\text{mix}\}\,,
\end{equation}
with 
\begin{align}
\label{E223} \g_T(\o,\vec{k}) &\equiv -C_T\frac{(\o^2-\vec{k}^2)^2}{48\r(\vec{k})}\bigg((M+\r(\vec{k}))\g_+(\o,\vec{k})(\o^2-\o_D(\vec{k})^2)-\\
\nn&- (M-\r(\vec{k}))\frac{\text{Im}\GG^-_R(\o,\vec{k})}{\text{Im}\GG^+_R(\o,\vec{k})} \g_-(\o,\vec{k})(\o^2-\o_s(\vec{k})^2)(\o^2-\o_s^
*(\vec{k})^2) \bigg) \, ,
\end{align}
and similar expressions for $\g_J$ and $\g_{\text{mix}}$.
In particular, in the extremal hydrodynamic regime $T\ll|\o|,|\vec{k}|\ll r_e^{-1}$, the IR correlators behave as $\text{Im}\GG^\pm_R(\o,\vec{k})\sim \o^{2\D_\pm(\vec{k})-1}$, from which we may infer the following properties:
\begin{itemize}
\item As expected \cite{Moitra:2020dal}, the leading order extremal hydrodynamic expressions
\begin{equation}
\label{E224a} \text{Im}G_T \propto \frac{(\o^2-\vec{k}^2)^2}{|d(\o,\vec{k})|^2}\o^{2\D_+(\vec{k})-1}(\o^2+\OO(r_e^4\vec{k}^4)) \, , 
\end{equation}
\begin{equation}
\label{E224b} \text{Im}G_J,\text{Im}G_{\text{mix}} \propto \frac{(\o^2-\vec{k}^2)\vec{k}^2}{|d(\o,\vec{k})|^2}\o^{2\D_+(\vec{k})-1}(\o^2+\OO(r_e^4\vec{k}^4))
\end{equation}
\begin{equation}
\nn d(\o,\vec{k}) \equiv (\o-\o_D(\vec{k}))(\o-\o_s(\vec{k}))(\o+\o_s^
*(\vec{k})) \, , 
\end{equation}
agree with the extrapolation of leading order hydrodynamics to zero temperature \cite{Kovtun:2012rj}, up to the modification of the low frequency power-law behavior from $\o$ to $\o^{2\D_+(\vec{k})-1}$;
\item The contribution of the sound channel correlator $G^+$ always dominates over that of the diffusive channel correlator $G^-$ for the stress-tensor and mixed correlators. For the current correlators, $G^-$ only dominates for $r_e|\vec{k}|\ll r_e^2\o^2$;
\item For the energy-energy, density-density and energy-density spectral functions, the height of the sound peak is of order $\OO((r_e|\vec{k}|)^{2\D_+(\vec{k})-3})$, whereas the diffusive peak is of order $\OO((r_e|\vec{k}|)^{4\D_+(\vec{k})-2})$. This indicates that the diffusive peak is suppressed at low temperatures, as was observed in \cite{Davison:2011uk} for $n=2$, and further provides the precise momentum scaling of this suppression in the extremal hydrodynamic regime. In particular, this scaling indicates that the diffusive peak tends to become more important as the momentum is increased, as was also observed in \cite{Davison:2011uk}.  
\end{itemize}

As a final remark, note that the ratio of the two IR correlators gives a power-law contribution $\o^{2(\D_+(\vec{k})-\D_-(\vec{k}))}$ to the generalized susceptibility \eqref{E223}. The latter is an example of a non-analytic term in the extremal hydrodynamic expansion, which resums a series of logarithmic terms 
\begin{equation}
\label{E225} (r_e\o)^{2(\D_+(\vec{k})-\D_-(\vec{k}))} = (r_e\o)^2\left(1+\frac{4 (3 n-4)}{3 n^2 (n+1)} (r_e\vec{k})^2 \log{(r_e\omega)+\OO(r_e^4\vec{k}^4)} \right) \, .
\end{equation}

\subsection{Magnetized branes}

\label{sec:mb}

For our last example, we consider a black brane background with $p=3$, meaning that in the near-extremal regime the theory flows to a CFT$_2$ in the IR. A solution of this sort was found in \cite{DHoker:2009mmn}, corresponding to a magnetized black brane at zero density. The action for the system is Einstein-Maxwell 
\begin{equation}
\label{M1} S = \int\intd x^{d+1}\sqrt{-g}\left(R + \frac{d(d-1)}{\ell^2} - \ell^2 F_{MN}F^{MN}\right) \, .
\end{equation}
where the boundary dimension $d\geq 4$ is taken to be even, and $\ell$ is the AdS length. The field strength ansatz is of the form 
\begin{equation}
\label{M2} F = B\left( \intd y_1\wedge\intd y_2 + \dots +\intd y_{d-3}\wedge\intd y_{d-2}\right) \, .
\end{equation}
with $B$ the constant boundary magnetic field. An appropriate ansatz for the metric is then given by
\begin{equation}
\label{M3} \intd s^2 = \ex^{2A(r)}\left(-f(r)\intd t^2+f(r)^{-1}\intd r^2+h(r)\intd x^2+\intd y_1^2+\dots+\intd y_{d-2}^2\right) \, ,
\end{equation}
with $r$ the radial coordinate, defined to be 0 at the AdS boundary, and $x$ the direction of the magnetic field. The ansatz for the field strength \eqref{M2} automatically solves the Maxwell equations, whereas the Einstein equations may be written in terms of the ansatz fields \eqref{M3} as
\begin{equation}
\label{M4} A''(r) - A'(r)\left(A'(r)+\frac{h'(r)}{2h(r)}\right) - \frac{f'(r)h'(r)}{2(d-1)f(r)h(r)} + \frac{2(B\ell)^2\ex^{-2A(r)}}{(d-1)f(r)} = 0 \, ,
\end{equation}
\begin{equation}
\label{M5} h''(r) + h'(r) \left((d-1)A'(r)+\frac{f'(r)}{f(r)}-\frac{h'(r)}{2 h(r)}\right)- \frac{4(B\ell)^2\ex^{-2A(r)}h(r)}{f(r)}  = 0 \, ,
\end{equation}
\begin{equation}
\label{M6} f(r) A'(r)\left(\frac{f'(r)}{f(r)}+\frac{h'(r)}{h(r)}+d A'(r) \right)+ \frac{f'(r)h'(r)}{2(d-1) h(r)}  -\frac{d\, \ex^{2 A(r)}}{\ell^2} + \frac{d-2}{d-1}\ex^{-2A(r)}(B\ell)^2 = 0 \, .
\end{equation}

Equations \eqref{M4}-\eqref{M6} admit a BTZ$\times\mathbb{R}^{d-2}$ solution \cite{DHoker:2009mmn}
\begin{equation}
\label{M7} \intd s^2 = \left(\frac{\ell_3}{\rho}\right)^2\left(-f_3(\rho)\intd t^2 + \frac{\intd\r^2}{f_3(\r)} + \intd x^2\right) + \frac{B\ell^2}{\sqrt{d-1}}\left(\intd y_1^2+\dots+\intd y_{d-2}^2\right) \, ,
\end{equation}
\begin{equation}
\label{M8} \r = \frac{1}{\sqrt{d-1}}\frac{1}{B(r_0-r)} \, ,
\end{equation}
where $r_0$ is an arbitrary integration constant, and the blackening function and AdS$_3$ length are given by
\begin{equation}
\label{M9} f_3(\r) = 1 - \left(\frac{\r}{\r_h}\right)^2  \quad ,\quad \ell_3 = \frac{\ell}{\sqrt{d-1}} \, ,
\end{equation}
with $\r_h$ the BTZ horizon radius. 

The near-extremal magnetic brane solution \cite{DHoker:2009mmn} interpolates between the BTZ solution \eqref{M7} near the horizon, and $AdS_{d+1}$ in the UV. For the fields of the ansatz \eqref{M3}, this means that when the boundary is approached ($r\to 0$) the fields behave as
\begin{equation}
\label{M10} A(r) = -\log{\left(\frac{r}{\ell}\right)} + \OO(r^d) \quad, \quad f(r) = 1 + \OO(r^d) \quad, \quad h(r) = 1 + \OO(r^d) \, ,
\end{equation}
whereas near the horizon ($r\to r_H$) we have
\begin{equation}
\label{M11a} A(r) = \frac{1}{2}\log{\left(\frac{B\ell^2}{\sqrt{d-1}}\right)} + \OO(r_H-r) \,,
\end{equation}
\begin{equation}
\label{M11b} f(r) = 4\pi T(r_H-r) + \sqrt{d-1}B(r_H-r)^2 + \OO(r_H-r)^3 \, ,
\end{equation}
\begin{equation}
\label{M11c} h(r) = \frac{4\pi^2 T^2}{\sqrt{d-1}B} + 4\pi T(r_H-r) + \sqrt{d-1}B(r_H-r)^2 + \OO(r_H-r)^3 \, .
\end{equation}

We now consider a probe neutral scalar field $\phi$ on the magnetic brane geometry, obeying the Klein-Gordon equation
\begin{equation}
\label{M12} \frac{1}{\sqrt{-g}}\pa_M\left(\sqrt{-g}\,g^{MN}\pa_N\phi\right) - m^2\phi = 0 \, .    
\end{equation}
The homogeneity of the background allows to expand $\phi$ in Fourier modes 
\begin{equation}
\label{M13} \phi(r;x) = \int\frac{\intd k^d}{(2\pi)^d}\ex^{-i\o t+ik_x x+ik_y y}\varphi_{\o,k_x,k_y}(r) \, ,
\end{equation}
where we took the momentum transverse to the magnetic field to lie in the $y \equiv y_1$ direction (without loss of generality). In Fourier space, \eqref{M12} becomes 
\begin{align}
\nn &\ex^{-(d-1)A(r)}h(r)^{-\frac{1}{2}}f(r)\pa_r\left(\ex^{(d-1)A(r)}h(r)^{\frac{1}{2}}f(r)\pa_r\varphi\right)+\\
\label{M14} &+ \left(\o^2-\frac{f(r)}{h(r)}k_x^2-f(r)k_y^2-\ex^{2A(r)}f(r)m^2\right)\varphi = 0 \, ,
\end{align}
where the indices were suppressed on $\varphi$ to avoid clutter. Introducing again the tortoise coordinates $\intd z/\intd r=1/f(r)$, \eqref{M14} may be put in Schr\"odinger form upon performing the field redefinition 
\begin{equation}
\label{M15} \varphi(z) \to \ex^{-\frac{d-1}{2}A(z)}h(z)^{-\frac{1}{4}} \psi(z) \, .
\end{equation}
The field $\psi$ obeys 
\begin{equation}
\label{M16} \pa_z^2\psi + (\o^2-V(z))\psi = 0 \, ,
\end{equation}
with the Schr\"odinger potential given by 
\begin{align}
\nn V(z) = &f(z)\left(\frac{k_x^2}{h(z)}+ k_y^2 + \ex^{2A(z)}m^2\right)+\\
\label{M17} &+ \frac{d-1}{2}A''(z) + \frac{d-1}{4}A'(z)\left((d-1)A'(z)+\frac{h'(z)}{h(z)}\right) - 3 \left(\frac{h'(z)}{4h(z)}\right)^2 + \frac{h''(z)}{4h(z)} \, .
\end{align}

The potential \eqref{M17} is regular everywhere and has UV asymptotics of the form \eqref{pf2d} with $\n^2 = (d-1)^2/4\geq 0$, which ensures that the scalar two-sided correlator $G_{12}(\o,k_x,k_y)$ associated with $\phi$ obeys the product formula without zeroes. The poles of the retarded correlator associated with $\phi$ can be computed from a numerical calculation, whose results are presented in appendix \ref{Sec::AppH}. These results indicate that $\phi$ does not feature any gapless pole, and its soft poles approach the IR CFT$_2$ poles at low temperature. According to the discussion of section \ref{sec:special}, the near-extremal limit of $G_{12}$ may then be written as 
\begin{equation}
\label{M19} G_{12}(\o,k_x,k_y) = \gamma_e(\o,k_x,k_y) \GG^{k_y}_{12}(\o,k_x)\left(1+\OO(r_e T)\right) \, ,
\end{equation}
with $\gamma_e$ analytic and free of zeroes, and $\GG_{12}^{k_y}(\o,k_x)$ the IR CFT$_2^{k_y}$ correlator associated with $\phi$. From the known expression for finite temperature CFT$_2$ two-point functions \cite{Dodelson:2023vrw,Son:2002sd}, $\GG_{12}$ may be written in terms of the IR conformal dimension $\D(k_y)$ as 
\begin{align}
\label{M19a} \GG_{12}^{k_y}(\o,k_x) \!= &\frac{(2\pi T)^{2\D(k_y)-2}}{\pi\Gamma\big(\D(k_y)-1\big)^2}\left|\Gamma\left(\frac{2\pi T\D(k_y)+i(\o+k_x)}{4\pi T}\right)\right|^2\!\left|\Gamma\left(\frac{2\pi T\D(k_y)+i(\o-k_x)}{4\pi T}\right)\right|^2 \!\!\! .
\end{align}
The dimension of the IR operator associated with $\phi$ can be read from the near-horizon limit of \eqref{M14}, which gives
\begin{equation}
\label{M21} \D(k_y) = 1 + \sqrt{1+ (m\ell_3)^2+\frac{k_y^2}{\sqrt{d-1}B}} \, .
\end{equation}

The extremal hydrodynamic product formula \eqref{neh4x} is obtained from the zero temperature limit of \eqref{M19a}
\begin{equation}
\label{M20} \text{Im}G_R(\o,k_x,k_y) =  \tilde{\g}_e(\o,k_x,k_y)(r_e^2(\o^2-k_x^2))^{\D(k_y)-1} \, .
\end{equation}
The zero-temperature effective susceptibility $\tilde{\g}_e$ is a function of $r_e^2(\o^2-k_x^2)$ and $r_e^2k_y^2$, which is free of zeroes and analytic in the regime $r_e^2(\o^2-k_x^2),r_e^2k_y^2\ll 1$.

\section{An application to holographic neutrino transport}

\label{sec:application}

In this last section, the extremal hydrodynamic product formula \eqref{neh4x} is applied to the setup considered in \cite{Jarvinen:2023xrx}. There, a toy model was introduced to investigate neutrino transport in holography, where the charged current spectral functions were computed from the Maxwell equations on the RN background, i.e. the problem considered in section \ref{sec:ccc}. The numerical data for the spectral functions was then used to compute the neutrino and anti-neutrino opacities, $\k$ and $\bar{\k}$. An interesting observation from this calculation was that the opacities at low temperatures were quite well approximated by substituting for the current correlators the leading order hydrodynamic expression \cite{Kovtun:2012rj}, extrapolated to zero temperature. 

As discussed in this work, the fact that the correlators can be approximated by a hydrodynamic expression is a direct consequence of the existence of a hydrodynamic-like gapless pole in the near-extremal regime (see appendix \ref{sec::AppB3}), due to the product formula. In section \ref{sec:ccc}, we further argued that the leading order extremal hydrodynamic product formula actually predicts expressions \eqref{E120} and \eqref{E123} which are slightly different from usual hydrodynamics. Specifically, the leading low-frequency behavior is changed from linear to $\o^{2\D(k)-1}$, where $\D(k)$ is the momentum-dependent scaling dimension in the IR CFT$_1$. This specific power-law is the main signature on the spectral functions form the IR CFT, and from the branch-cut that arises in the zero-temperature limit. Here, we investigate how this improved description of the spectral functions affects the calculation of the neutrino opacities. 

\begin{figure}[h]
\begin{center}
\includegraphics[scale=0.5]{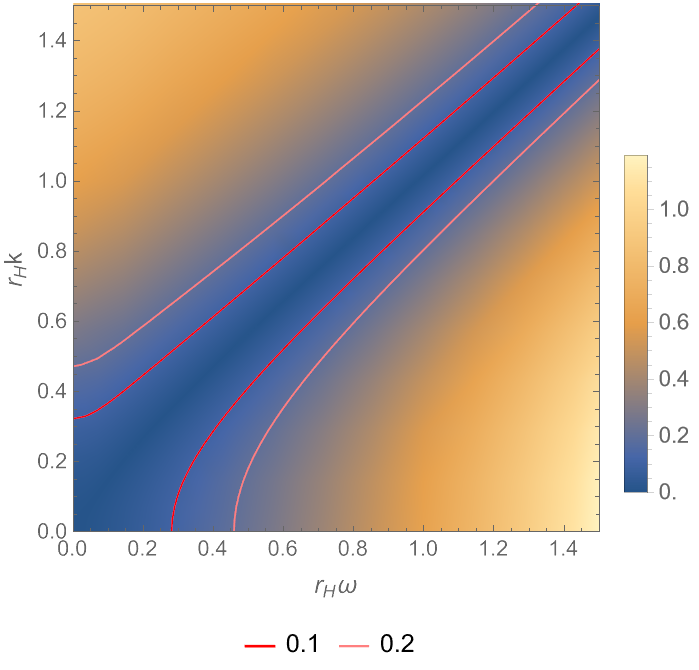}
\hfill
\includegraphics[scale=0.5]{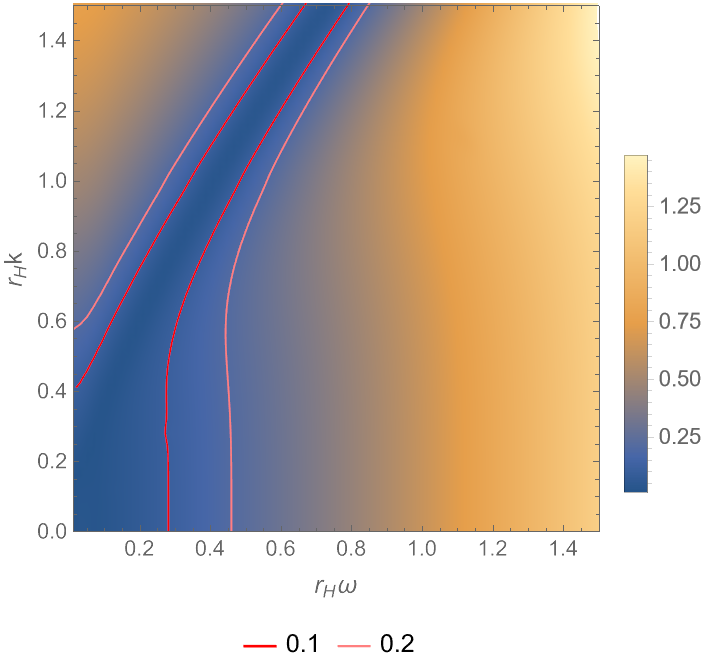}
\includegraphics[scale=0.5]{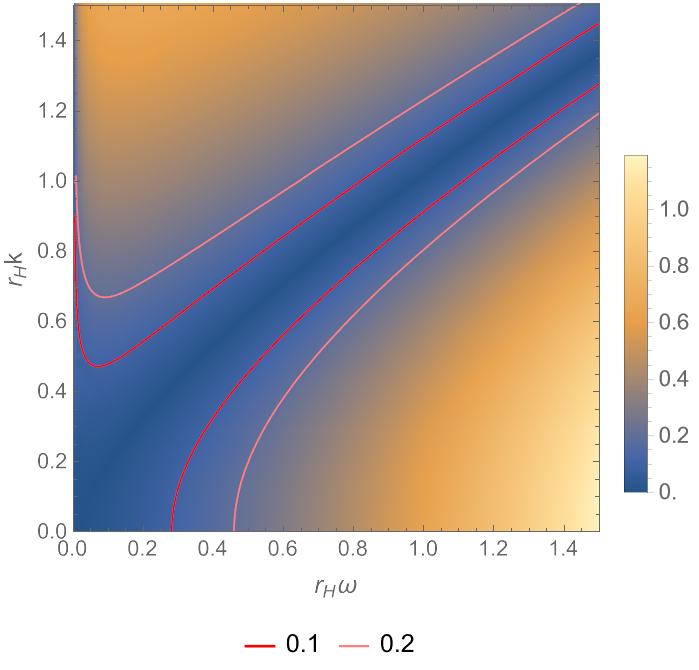}
\hfill
\includegraphics[scale=0.5]{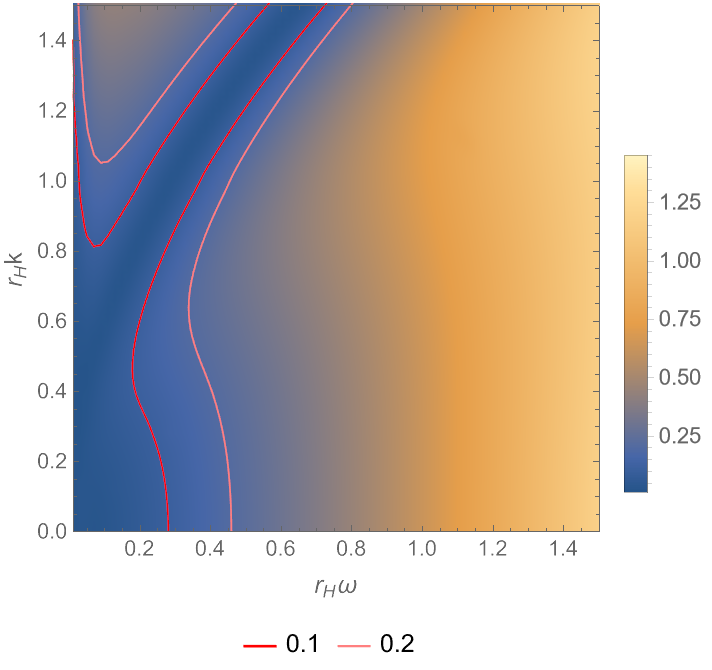}
\caption{Relative difference of the numerically computed transverse (left) and longitudinal (right) spectral functions, with the leading order hydrodynamic (top) and extremal hydrodynamic (bottom) approximations. The relative differences are shown as a function of frequency $\o$ and momentum $|\vec{k}|$, in units of the horizon radius $r_H$. The chemical potential to temeperature ratio is $\mu/T\simeq 65$.}
\label{Fig:RelDiff}
\end{center}
\end{figure}

We first discuss how the change in the hydrodynamic expressions affects the approximation to the spectral functions themselves. This is presented in figure \ref{Fig:RelDiff}, which shows a comparison between, on the one hand, the relative difference of the exact (numerial) spectral functions with standard hydrodynamics (top), and with the extremal hydrodynamic expressions \eqref{E120} and \eqref{E123} (bottom), at $\m/T \simeq 65$. The conductivity $\sigma = r_H^{-1}/(2\pi^2)$ was chosen as in \cite{Jarvinen:2023xrx}. As expected (see section \ref{sec:pfne}), these plots indicate that the main effect of taking into account the IR dimension is to improve the description of the spectral functions at low frequency $\o$.    

\begin{figure}[h]
\begin{center}
\includegraphics[scale=0.8]{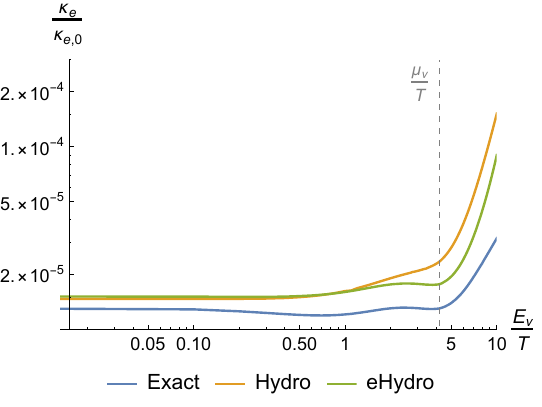}
\hfill
\includegraphics[scale=0.8]{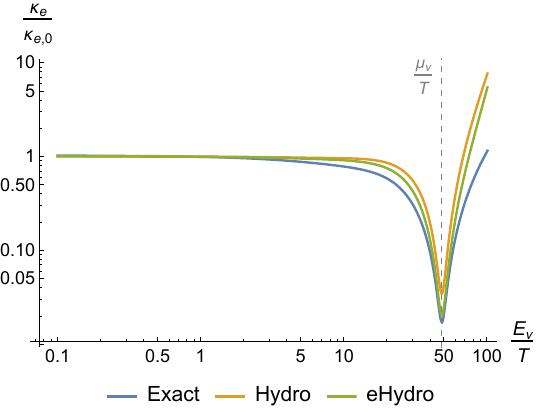}
\caption{Neutrino opacity $\k$ and its hydrodynamic approximations as a function of the
neutrino energy $E_\n$, for $\m/T\simeq4.7$ (left) and $\m/T\simeq 65$ (right). $\k_{e,0}$ is a normalization factor that was introduced in \cite{Jarvinen:2023xrx} and $\mu_\n$ the neutrino chemical potential, whose location is indicated by the gray dashed lines.}
\label{Fig:kaenu}
\end{center}
\end{figure}
\begin{figure}[h]
\begin{center}
\includegraphics[scale=0.8]{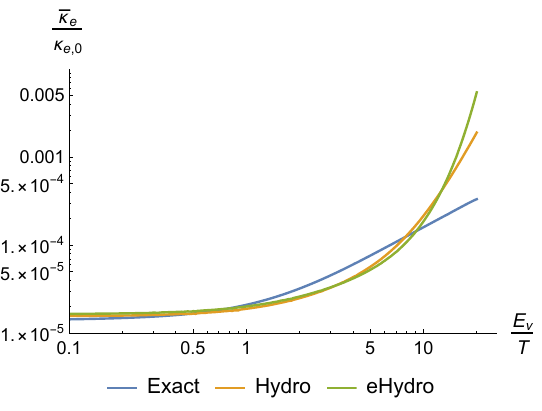}
\hfill
\includegraphics[scale=0.8]{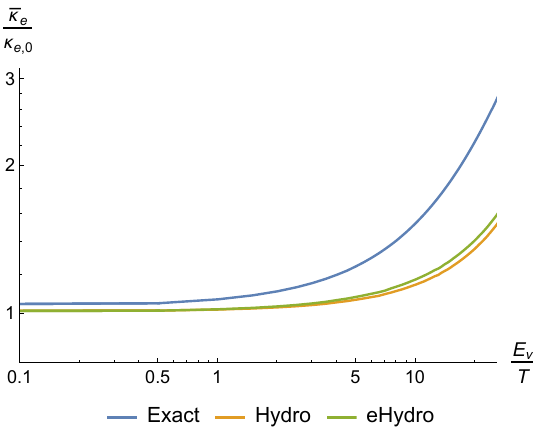}
\caption{Same as figure \ref{Fig:kaenu} but for the anti-neutrino opacity $\bar{\k}$.}
\label{Fig:kaenub}
\end{center}
\end{figure}

The better description of the correlators at low energies translates into a better approximation to the neutrino opacities, as illustrated in figures \ref{Fig:kaenu} and \ref{Fig:kaenub}. The latter show a comparison of the exact holographic opacities with the hydrodynamic and extremal hydrodynamic approximations, as a function of neutrino energy $E_\n$ and for two different values of the ratio of chemical potential to temperature\footnote{For $T=10\,\text{MeV}$, these values correspond to baryon number densities $n_{B,1}=10^{-3\,}\mathrm{fm^{-3}}$ and $n_{B,2}=1\,\mathrm{fm^{-3}}$, which covers the typical range expected in neutron stars \cite{Jarvinen:2023xrx}.}: $\mu_1/T\simeq 4.7$ and $\mu_2/T\simeq 65$. For most values of $E_\n$, the extremal and standard hydrodynamic approximations are observed to have similar accuracies, except near the dip in the neutrino opacity around $E_\n\sim\m_\n$ (with $\mu_\n$ the neutrino chemical potential; see figure \ref{Fig:kaenu}), where the extremal hydrodynamic approximation gives a significantly better description. This is consistent with expectations, since the integrand from which the neutrino opacity is computed (see \cite{Jarvinen:2023xrx}) is concentrated near $\o=0$ in this case. Note that this regime of neutrino energies is particularly important to describe well since it corresponds to the effective Fermi surface\footnote{See \cite{Jarvinen:2023xrx} for a discussion of the quasi-particle approximation considered for the description of neutrinos.}, near which most neutrinos are emitted and absorbed.

\section{Discussion}

\label{discussion}

In this work, we used the holographic product formula introduced in \cite{Dodelson:2023vrw} to get insight into the structure of holographic spectral functions in the near-extremal hydrodynamic regime, $\o,k,T\ll\m$. In the limit of vanishing temperature $T\ll\o,k\ll \m$, the formula provides an expression where both the low temperature gapless modes and the IR scaling behavior appear explicitly, moreover in a simple factorized form. As we showed in sections \ref{sec:Ex} and \ref{sec:application}, the product formula applies in many holographic settings, where it allows to extend the near-extremal hydrodynamic approximation of correlators to the regime of low energies.

Several directions may be envisaged for future developments on this topic:
\begin{itemize}
\item Here we mostly focused on the pole structure, but also found in the various examples that the properties of the Schr\"odinger potential provide information on the zeroes of the spectral functions. It would be interesting to study further the relation between zeroes and singularities of the potential, especially since zeroes are typically much harder to compute than poles in holography;    

\item Another interesting point is related to the recent results of \cite{Grozdanov:2025ner,Grozdanov:2025ulc} in four bulk dimensions, which showed that duality relations between different parity sectors of correlators translate into additional constraints on their structure. These may allow to write even more useful expressions in the near-extremal hydrodynamic regime of $(2+1)$-dimensional holographic theories;  

\item In \cite{Dodelson:2023vrw}, the UV limit $\o\gg T$ of the product formula was analyzed in relation with the OPE in the UV CFT. Here, we were rather interested in an IR regime of the correlators $\o \ll \m$, but also considered the limit $T\ll\o\ll\m$ to write the extremal hydrodynamic product formula. In this case, the corrections of order $\OO(T/\o)$ to the extremal result may still be controlled by an OPE, although rather associated with the IR CFT, which could be worth investigating;

\item We focused on situations where the bulk fluctuation equations can be decoupled, especially since the IR correlator emerges simply in this case. Even when the equations cannot be decoupled, if the background geometry asymptotes to AdS$_p$ near the horizon, then the low energy limit of correlators should still be controlled by the IR CFT. It would be nice to understand how this comes about in the more general coupled case, and generalize the near-extremal product formulae accordingly.
\end{itemize}

Finally, the main problem remains to understand what is the structure of the low energy effective field theory (EFT) that underlies the near-extremal spectral functions, which should combine both hydrodynamic-like features and the IR CFT. The formalism of \cite{Crossley:2015evo,Glorioso:2017fpd,Liu:2018kfw} should a priori encompass this kind of theories, since it is well adapted to describe systems with hydrodynamic-like features \cite{Blake:2017ris}. This type of effective theories are also known to arise in holographic calculations \cite{Nickel:2010pr,Glorioso:2018mmw,Davison:2022vqh}, which however still remain to be extended to the relevant systems.

\section*{Acknowledgements}
I would like to thank Andrew Gomes and Eren Firat for interesting discussions on the topic of this work. I am also grateful to Blaise Gout\'eraux and Jay Armas for useful comments on a draft. This project has received funding from the European Union’s Horizon 2024 research and innovation program under the Marie Sklodowska-Curie grant agreement No 101210184.

\clearpage
\appendix

\section{IR limit of near-extremal correlators}
\label{sec::AppA}

In this appendix, we show how the near-extremal correlators computed from the Schr\"odinger equation
\begin{equation}
\label{A1} \psi''(z) + (\o^2-V(z)) \psi(z) = 0 \, ,
\end{equation}
are related in the IR limit to correlators in the IR CFT$_{p-1}$. For general $p$, the IR limit is defined by 
\begin{equation}
\label{A2} k_{p-1}^2 = -\o^2 + \vec{k}_{p-2}^2 \to 0 \, .
\end{equation}
Since the background is near-extremal, the temperature $T$ is also much smaller than the scale $\m$, so the limit depends on the hierarchy between $|k_{p-1}|$ and $T$. We will consider the most general case where the two are of the same order\footnote{Other cases can be obtained by taking appropriate limits of the result with $|k_{p-1}|$ and $T$ of the same order.}
\begin{equation}
\label{A3} k_{p-1}^2 \to \e^2 k_{p-1}^2 \quad,\quad T \to \e T \quad,\quad \e \ll 1 \, .
\end{equation}
Since the system becomes Lorentz invariant along the directions of the IR CFT at zero temperature, it is convenient to redefine the Schr\"odinger potential such that \eqref{A1} takes the form
\begin{equation}
\label{A4} \psi''(z) + (\o^2-\vec{k}_{p-2}^2-V(z))\psi(z) = 0 \, .
\end{equation}

Now, the procedure to exhibit the relation with the IR correlator is based on the standard division of the bulk into inner and outer region \cite{Faulkner:2009wj,DHoker:2010xwl}:
\begin{itemize}
\item The inner region is where $z\gg 1$. Taking 
\begin{equation}
\label{A5} z \equiv \e^{-1} \zeta \, ,
\end{equation}
with $\e$ the same small parameter that appears in \eqref{A3} and $\zeta$ of order 1, the near-extremal backgrounds of interest are such that the potential in  \eqref{A4} behaves as 
\begin{equation}
\label{A6} V(z) \sim \e^2 \zeta^{-2}V_p(\zeta T,\zeta^2k_{p-1}^2,r_e^2\vec{k}_{d-p+1}^2) \, .
\end{equation}
The Schr\"odinger equation in the inner region then takes the form
\begin{equation}
\label{A7} \psi''(\zeta) - (k_{p-1}^2+\zeta^{-2}V_p(\zeta)) \psi(\zeta) = 0 \, , 
\end{equation}
which should be the equation of motion for a fluctuation in AdS$_p$-Schwarzchild. In particular, near the AdS$_p$ boundary ($\zeta\to 0$), the dependence of $V_p$ on the temperature and CFT momentum drops, so that \eqref{A7} takes a homogeneous form 
\begin{equation}
\label{A8} \psi''(\zeta) - \frac{\a(r_e^2\vec{k}_{d-p+1}^2)}{\zeta^2}\psi(\zeta) = 0 \, , 
\end{equation}
which has two independent solutions 
\begin{equation}
\label{A9} \psi_\pm(\zeta) = \zeta^{\frac{1}{2}\pm\n(r_e^2\vec{k}_{d-p+1}^2)} \quad,\quad \n(r_e^2\vec{k}_{d-p+1}^2) \equiv \frac{1}{2}\sqrt{1+4\a(r_e^2\vec{k}_{d-p+1}^2)} \, .
\end{equation}
By fixing the normalization of $\psi(z)$, we may therefore write the solution near the boundary of the inner region as 
\begin{equation}
\label{A10} \psi(\zeta) = \zeta^{\frac{1}{2}-\n}(1+\dots) + \GG(\o,\vec{k})\zeta^{\frac{1}{2}+\n}(1+\dots) \, ,
\end{equation}
where the coefficient of the subleading term is the IR correlator by definition \cite{Faulkner:2009wj}, and the dots go to zero as $\zeta\to 0$. Note that $\D_{\text{IR}} = 1/2+\n$ is (half) the scaling dimension of the IR correlator associated with $\psi$.   

\item The outer region is where $\zeta\ll 1$. In particular, it overlaps with the inner region near the boundary of the IR AdS$_p$, where the outer solution behaves as in \eqref{A10}. By linearity of \eqref{A4}, the full outer solution may then be written as \cite{Faulkner:2009wj}
\begin{equation}
\label{A11} \psi_{\text{out}}(z) = \eta_-(z) + \GG(\o,\vec{k}) \eta_+(z) \, .
\end{equation}
Near the UV boundary ($z\to 0$), $\eta_\pm(z)$ obeys an expansion controlled by (half) the dimension $\D_{\text{UV}}$ of the UV correlator associated with $\psi$, which takes the form
\begin{equation}
\label{A12} \eta_\pm(z) = a_\pm(\o,\vec{k}) z^{1-\D_{\text{UV}}}(1+\dots) + b_\pm(\o,\vec{k}) z^{\D_{\text{UV}}}(1+\dots) \, , 
\end{equation}
where the dots vanish at $z=0$. The correlator associated with $\psi$ may then be expressed as 
\begin{equation}
\label{A13} G_\psi(\o,\vec{k}) = \frac{b_-(\o,\vec{k}) + b_+(\o,\vec{k})\GG(\o,\vec{k})}{a_-(\o,\vec{k}) + a_+(\o,\vec{k})\GG(\o,\vec{k})} \, .  
\end{equation}
Equation \eqref{neh4c} in the main text is obtained from the expansion of the imaginary part\footnote{Taking into account that $\eta_\pm$ are real for real $\o$ and $\vec{k}$.} of \eqref{A13} at small $r_e|\o|$ and $r_e|\vec{k}|$, in the case where $G$ is the retarded correlator (i.e. for $\psi$ obeying infalling boundary conditions at the horizon).  

\end{itemize}

\section{An example at next-to-leading order in the near-extremal hydrodynamic expansion}
\label{sec::AppB}

We discuss here how the correlator computed in \cite{Gouteraux:2025kta} at next-to-leading order in the near-extremal hydrodynamic expansion fits with the product formula \eqref{neh4}, and in particular with its extremal limit \eqref{neh4x}. 

The correlator in question corresponds to the longitudinal retarded two-point function for a U(1) current in the neutral AdS$_4$-black-brane state with metric\footnote{Note that this is the same problem as in section \ref{sec:ccc}, only for a specific choice of the blackening function $f(r)$.} 
\begin{equation}
\nn \intd s^2 = \frac{\ell^2}{r^2}\left( -f(r)\intd t^2+f(r)^{-1}\intd r^2+\intd\vec{x}^2 \right) \, ,
\end{equation}
\begin{equation}
\label{B1} f(r) = 1 - \frac{m^2 r^2}{2} - \left(1-\frac{m^2r_H^2}{2}\right)\frac{r^3}{r_H^3} \, ,
\end{equation}
where $\ell$ is the AdS length, $r_H$ the horizon radius and $m$ the source for the two massless scalar fields that are also turned on in the background 
\begin{equation}
\label{B2} \psi_i = m x^i \, , \quad i\in\{1,2\} \, .
\end{equation}
In the regime 
\begin{equation}
\label{B2b} |\o|\sim|\vec{k}^2|/m\sim T = \OO(\e) m \, , \quad \e \ll1 \, ,
\end{equation}
the result of \cite{Gouteraux:2025kta} takes the form\footnote{Note that we also need $i\o/(2\pi T) \notin \mathbb{N}^*$ for $\GG$ in \eqref{B4} to be finite. In other words, the expansion \eqref{B3} breaks down when $i\o/(2\pi T)$ is a positive integer.}
\begin{equation}
\label{B3} G^R_{xx}(\o,\vec{k}) = \frac{r_e^{-1} \bm{\o}^2\left(1-\frac{1}{3}\bm{\vec{k}}^2\GG\right) + \OO(\e^4)}{i\bm{\o}- \bm{\vec{k}}^2+ \frac{1}{6}\bm{\vec{k}}^2\left(4\pi\bm{T} + 2(\bm{\vec{k}}^2+i\bm{\o})\GG + 3\bm{\vec{k}}^2 \log{3}\right)+\OO(\e^3)} \, ,
\end{equation}
\begin{equation}
\label{B4} \GG \equiv \pi \cot{\left(\frac{i\o}{2T}\right)} + \gamma + \psi{\left(\frac{i\o}{2\pi T}\right)} - \log{\left(\frac{9}{4\pi\bm{T}}\right)} \, ,
\end{equation}
with $x$ the direction of the momentum $\vec{k}$, $r_e \sim m^{-1}$ the extremal horizon radius and bold fonts indicating quantities measured in units of $r_e^{-1}$, e.g. $\bm{\omega}\equiv r_e\o$. In \eqref{B4}, $\psi$ is the digamma function and $\g$ the Euler constant.

We are interested in the limit\footnote{Note that, in this case, the soft poles do not match the poles of the IR CFT correlator as in \eqref{neh4f} \cite{Gouteraux:2025kta}. There is therefore no simple expression of the form \eqref{neh6} for the correlator in the general near-extremal regime $|\o|\sim|\vec{k}^2|\sim T\ll r_e^{-1}$. However, the extremal formula \eqref{neh4x} still applies.} $\o/T\to\infty$, for which the product formula \eqref{neh4} simplifies according to \eqref{neh4e}, whereas \eqref{B3}-\eqref{B4} becomes (for $\o$ off the negative imaginary axis)
\begin{equation}
\label{B5} G^R_{xx}(\o,\vec{k}) = \frac{r_e^{-1} \bm{\o}^2\left(1-\frac{1}{3}\bm{\vec{k}}^2\GG_0\right)+ \OO(\e^4)}{i\bm{\o}- \bm{\vec{k}}^2+ \frac{1}{6}\bm{\vec{k}}^2\left(2(\bm{\vec{k}}^2+i\bm{\o})\GG_0 + 3\bm{\vec{k}}^2 \log{3}\right)+ \OO(\e^3)} \, ,
\end{equation}
\begin{equation}
\label{B5b} \GG_0 \equiv \g - i\frac{\pi}{2} + \log{\left(\frac{2\bm{\o}}{9}\right)} \, .
\end{equation}
At this order equation \eqref{B5} has two poles \cite{Gouteraux:2025kta}
\begin{equation}
\label{B6} \bm{\o}_\pm(\vec{k}) = -i \bm{\vec{k}}^2 + i\frac{2}{3}\bm{\vec{k}}^4\left(\log{(2\bm{\vec{k}}^2)}+\g-\frac{5}{4}\log{3} \mp i\pi\right) + \OO(\bm{\vec{k}}^6) \, .
\end{equation}
This result indicates that the leading order diffusive pole $\bm{\o} = -i\bm{\vec{k}}^2$ splits into two at next-to-leading order. 

The imaginary part of \eqref{B5} can now be written in the form of the product formula \eqref{neh4}
\begin{equation}
\label{B7} \text{Im}G^R_{xx}(\o,\vec{k}) =  \frac{r_e^{-1}g(\bm{\o},\bm{\vec{k}})}{\left(\bm{\omega}^2-\bm{\omega}_+(\vec{k})^2\right)\left(\bm{\o}^2-\bm{\omega}_-(\vec{k})^2\right)} \, , 
\end{equation}
with the numerator given by
\begin{align}
\nn g(\bm{\o},\bm{\vec{k}}) = -\bm{\omega}  \Bigg[\bm{\bm{\omega}^2+\vec{k}}^4 - \frac{1}{3} \bm{\vec{k}}^2 \Bigg(&2 \left(\bm{\omega}^2-\bm{\vec{k}}^4\right) \log{\left(\frac{2 \bm{\omega} }{9}\right)}+ 8 \bm{\vec{k}}^4 \log \left(\frac{2 \bm{\vec{k}}^2}{9}\right) +  \\
\label{B8} & + 2 \gamma \bm{\omega}^2 + \bm{\vec{k}}^4 (6 \gamma +3 \log (3))- 2 \pi \bm{\vec{k}}^2 \bm{\omega} \Bigg)\Bigg] +\OO(\e^5) \, .
\end{align}
Finally, to recover \eqref{neh4x}, we factor out the IR power-law in \eqref{B8} as
\begin{equation}
\label{B9} g(\bm{\o},\bm{\vec{k}}) \equiv \bm{\o}^{2\D(\bm{\vec{k}})-1} \g_e(\bm{\o},\bm{\vec{k}}) \, ,
\end{equation}
with the IR scaling dimension given in this case by \cite{Gouteraux:2025kta}
\begin{equation}
\label{B10} \D(\bm{\vec{k}}) = \frac{1}{2}+\frac{1}{2}\sqrt{1+\frac{4}{3}\bm{\vec{k}}^2} \, ,
\end{equation}
and the generalized susceptibility $\g_e$ identified as 
\begin{align}
\nn \g_e(\bm{\o},\bm{\vec{k}}) = -(\bm{\omega}^2+\bm{\vec{k}}^4)+\frac{1}{3} \bm{\vec{k}}^2 \Bigg(&4 \bm{\omega} ^2 \log (\bm{\omega} )+ 8 \bm{\vec{k}}^4 \log \left(\bm{\vec{k}}^2\right)+ 2 \bm{\omega}^2 \left(\gamma -\log \left(\frac{9}{2}\right)\right)+\\
\label{B11} &+\bm{\vec{k}}^4 (6 \gamma -9 \log (3)+6\log (2))-2 \pi  \bm{\vec{k}}^2 \bm{\omega} \Bigg) + \OO(\e^4) \, .
\end{align}
In particular, the $\bm{\vec{k}}^6\log{\bm{\o}}$ term from \eqref{B8} is seen to have been absorbed into the IR power-law factor, such that at least at this order the extremal hydrodynamic expansion of $\g_e$ remains valid in the limit of vanishing $\bm{\o}$. It is plausible that the surviving $\bm{\o}^2\bm{\vec{k}}^2\log{(\bm{\o})}$ term could also be resummed into a power-law involving $\D(\bm{\vec{k}})$, such as $\bm{\o}^{2\D(\bm{\vec{k}})}$ or $\bm{\o}^{\D(\bm{\vec{k}})+1}$. 

Another special feature of \eqref{B11} is the existence of a zero at $\o = \vec{k}^2 = 0$. As mentioned in the main text, this zero has to be there to account for the fact that the two gapless poles \eqref{B6} merge into a single diffusive pole at leading order in the hydrodynamic expansion. At leading order, the product formula \eqref{neh4x} indeed predicts
\begin{equation}
\label{B12} \text{Im}G^R_{xx}(\o,\vec{k}) =  \frac{r_e^{-1}\bm{\o}^{2\D(\bm{\vec{k}})-1}\g_e(0)}{\bm{\omega}^2+\bm{\vec{k}}^4} \, , 
\end{equation}
which can be checked to be implied from \eqref{B7}-\eqref{B11} at leading order, with $\g_e(0) = -1$.

\section{Correlators for different field variables in the probe longitudinal gauge sector}

\label{sec::AppB2}

In this appendix, we demonstrate the relation \eqref{E116} for the correlators $\Pi^\parallel$ and $G_\Psi$, associated respectively to the longitudinal electric field $E^\parallel$ and the alternative field variable 
\begin{equation}
\label{cv1} \Psi(r) = \frac{f(r)\pa_rE^\parallel}{r^{(d-3)/2}(\o^2-f(r)k^2)} \, . 
\end{equation}

The correlators may be extracted from the near-boundary $(r\to0)$ expansion of the fields, which takes the form 
\begin{align}
\nn E^\parallel(r) = E_{(0)}&\left(1-\frac{\o^2-\vec{k}^2}{2(d-4)}r^2+\OO(r^4)\right)+\\
\label{cv2} & + r^{d-2}\left(- a_d(\o^2-\vec{k}^2)^{\frac{d-2}{2}} E_{(0)}\log{r} + E_{(2)}(1 + \OO(r^2))\right) \, ,
\end{align}
\begin{align}
\nn \Psi(r) = \frac{1}{4-d}&E_{(0)}r^{\frac{5-d}{2}}(1+\OO(r^2))+\\
\label{cv3} &+(d-2) r^{\frac{d-3}{2}} \left(-a_d E_{(0)} \log (r) \left(\omega^2-\vec{k}^2\right)^{\frac{d-4}{2}}+ \frac{E_{(2)}}{\omega^2 - \vec{k}^2}(1+\OO(r^2))\right) \, ,
\end{align}
with $a_d$ a dimension-dependent number, which vanishes for odd $d$. Up to a normalization constant, the retarded polarization function is then defined as  
\begin{equation}
\label{cv4} \Pi^\parallel_R = \frac{E_{(2)}}{E_{(0)}} \, ,
\end{equation}
for infalling boundary conditions at the horizon. 

On the other hand, from \eqref{cv3}, the retarded correlator $G^R_\Psi$ associated to the $\Psi$ variable is given by  
\begin{equation}
\label{cv7} G_\Psi^R = \frac{E_{(2)}}{(\o^2-\vec{k}^2)E_{(0)}} = \frac{\Pi^\parallel_R}{\o^2-\vec{k}^2} \, ,
\end{equation}
again up to a normalization constant. We therefore conclude that
\begin{equation}
\label{cv8} G_\Psi \propto \frac{\Pi^\parallel_{12}}{\o^2-\vec{k}^2} \, ,
\end{equation}
is free of zeroes.

\section{Poles of the probe current correlators}

\label{sec::AppB3}

In this appendix, we present a numerical calculation of the low-lying (gapless and soft) poles of the RN probe current correlators discussed in section \ref{sec:ccc}, in the near-extremal regime. We used for this the Mathematica package of \cite{Jansen:2017oag}. The transverse and longitudinal sectors are discussed separately. 

\subsection{Transverse correlator}

\begin{figure}[h]
\begin{center}
\includegraphics[scale=1.]{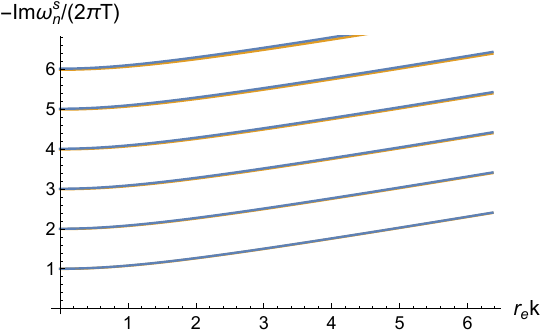}
\caption{Imaginary part of the first soft poles of the transverse polarization function in units of $2\pi T$ and as a function of momentum $r_e|\vec{k}|$, for $d=4$ and $r_e T = (50\pi)^{-1}$. The blue dots are the numerical results whereas the orange lines show the poles of the IR CFT$_1$ correlator \eqref{tc1}.}
\label{Fig:poles_Pit}
\end{center}
\end{figure}

The transverse sector does not feature any gapless pole, and the soft poles are purely imaginary. The momentum dependence of the first few soft poles is shown in figure \ref{Fig:poles_Pit}, for $d=4$ and $r_eT = (50\pi)^{-1} \simeq 6.4\times 10^{-3}$. This shows a very good agreement with the IR poles\footnote{The expression for the IR conformal dimension $\D(\vec{k})$ is obtained by substituting the RN blackening function \eqref{E20b} into \eqref{E118}.} (shown as the orange lines)
\begin{equation}
\label{tc1} \o_n^{IR}(\vec{k}) = -i2\pi T(\D(\vec{k})+n) \, ,\quad \D(\vec{k}) = \frac{1}{2}+ \frac{1}{2} \sqrt{1+\frac{4}{d(d-1)}(r_e\vec{k})^2} \, ,
\end{equation}
up to small corrections that can be checked to be of order $\OO(r_e T)$. 

\subsection{Longitudinal correlator}

\begin{figure}[h]
\begin{center}
\includegraphics[scale=1.]{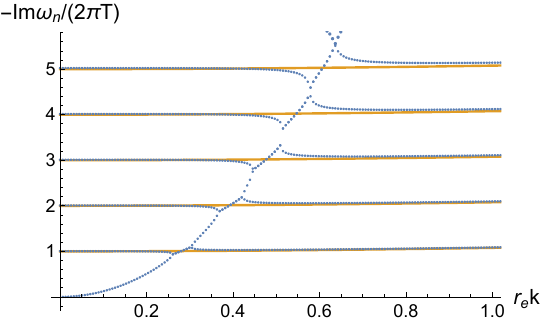}
\caption{Imaginary part of the first low-lying poles of the longitudinal polarization function in units of $2\pi T$ and as a function of momentum $r_e|\vec{k}|$, for $d=4$ and $r_e T = (50\pi)^{-1}$. The blue dots are the numerical results whereas the orange lines show the poles of the IR CFT$_1$ correlator \eqref{tc1}.}
\label{Fig:Impoles_Pil}
\end{center}
\end{figure}

\begin{figure}[h]
\begin{center}
\includegraphics[scale=1.]{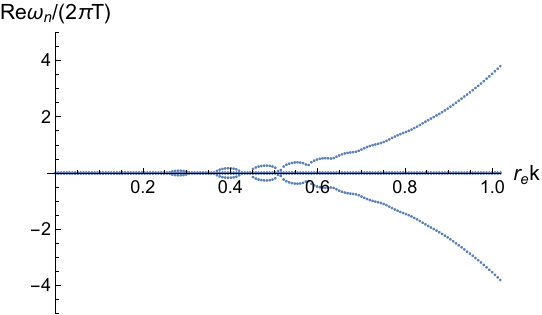}
\caption{Real part of the poles shown in figure \ref{Fig:Impoles_Pil}. For every momentum, at most two poles have finite real parts.}
\label{Fig:Repoles_Pil}
\end{center}
\end{figure}

We now present the numerical results for the poles of the longitudinal correlator, which are shown in figures \ref{Fig:Impoles_Pil} (imaginary part) and \ref{Fig:Repoles_Pil} (real part), still with $d=4$ and $r_e T = (50\pi)^{-1}$. This sector contains both a diffusive pole $\o_D(\vec{k}) = -iD\vec{k}^2 + \OO(\vec{k}^4)$ and a sequence of soft poles, which interact non-trivially. The behavior as momentum increases is very similar to \cite{Gouteraux:2025kta}, with the diffusive pole splitting into two poles with finite real parts before recombining on the imaginary axis several times as it collides with soft poles, before eventually remaining outside the imaginary axis, with a behavior of the form \eqref{B6}. Figure \ref{Fig:Impoles_Pil} indicates that the soft poles asymptote to the IR poles \eqref{tc1} (orange lines) at large momenta, but deviate significantly around the point where they cross the diffusive mode.  

\section{Coefficients of the helicity-0 fluctuation equations}

\label{sec::AppC}

The coefficients appearing in \eqref{E22}-\eqref{E22b} are given by
\begin{equation}
\label{E22c} \zeta(r) \equiv r f'(r)-(n+1) (f(r)-1) \, ,
\end{equation}
\begin{align}
\nn P_Z(r) \equiv & -2\vec{k}^2r^2 \left(rf'(r) + (n-2)f(r)\right) + n rf'(r)\left(rf'(r)-5nf(r)\right)+\\
\label{E22d}&+ 4n^2(n+1) f(r)(f(r)-1) \, ,
\end{align} 
\begin{align}
\nn P_S(r) \equiv &\,\,4 \vec{k}^2 r^2 \left(n(n-2)f(r)+\vec{k}^2 r^2\right) - n^2 r f'(r) \left(r f'(r)-2(2n+1)f(r) \right)-\\
\label{E22e} &- 4n^3(n+1)f(r)(f(r)-1) \, ,
\end{align}
\begin{align}
\nn V_S(r) = \frac{f(r)}{16r^2H(r)^2} \Big\{&4 \vec{k}^2 r^2 \Big(4 \vec{k}^4 r^4+\\ 
\nn&\quad\qquad + \vec{k}^2 r^2 \left(2 (5 n-4) r f'(r)+(8-7 n (n+2)) f(r)+8 n (n+1)\right) \!+ \\
\nn &\quad\qquad+n^2 (19 n-30) r f'(r) f(r)-n(n-4) r^2 f'(r)^2-\\
\nn &\quad\qquad -8 n^2 (n+1) (2 n-3) f(r)(f(r)-1)\Big) + \\
\nn &+n^3 r f'(r) \Big(((41 n+26) f(r)-8 (n+1)) rf'(r)-10 r^2 f'(r)^2-\\
\nn &\qquad\qquad-16 (n+1) (4 n+1) f(r)(f(r)-1) \Big)+\\
\label{E22f} &+32 (n+1)^2 n^4 (f(r)-1)^2 f(r)\Big\} \, .
\end{align}

\section{Stress-tensor correlators for other numbers of dimensions}
\label{sec::AppD}

In this appendix, we discuss the problem of section \ref{sec:sound} for other numbers of dimensions, $n=2$ and $n\geq4$. We first review the case of $n=2$, before considering $n\geq 4$.

\subsection{\texorpdfstring{$\bm{n=2}$}{n=2}}

The helicity-0 correlators on the AdS$_4$-RN background were studied in \cite{Edalati:2010pn}. As for $n=3$, the system of fluctuation equations can be  decoupled by considering the Kodama-Ishibashi (KI) master field $\Phi_\pm$ in \eqref{E23}. However, for $n=2$, the relation between the near-boundary expansions of $\Phi_\pm$
\begin{equation}
\label{n21} \Phi_\pm(r) = \Phi_\pm^{(0)} (1+\OO(r^2)) + \Phi_\pm^{(1)}r(1+\OO(r^2)) \, ,
\end{equation}
and the original metric and gauge field variables
\begin{equation}
\label{n22} \d g^\m_{\hp{\m}\n} = \d g^{(0)\m}_{\hp{(0)\m}\n}(1+\OO(r^2)) + \d\pi^{\m}_{\hp{\m}\n} r^3(1+\OO(r^2))  \, ,
\end{equation}
\begin{equation}
\label{n23} \d A_\m = \d A_\m^{(0)}(1+\OO(r^2)) + \d\pi_\m r(1+\OO(r^2)) \, ,
\end{equation}
is such that the leading and subleading coefficients of $\Phi_\pm$ both contain a mixture of sources and vevs for the boundary operators. Explicitly \cite{Edalati:2010pn} 
\begin{equation}
\label{n24} \Phi_\pm^{(0)} = \frac{1}{\bm{\vec{k}}^2}\left(-3r_H^{2}\a_{\pm,0}\d\pi^t_{\hp{t}t}-\bm{\vec{k}}^2\d\pi_t + 4r_H^2 Q\a_{\pm,0}^2\d g^{(0)y}_{\hp{(0)y}y} - \frac{1}{2}Q\bm{\vec{k}}^2\d g^{(0)t}_{\hp{(0)t}t}\right) \, ,
\end{equation}
\begin{align}
\nn \Phi_\pm^{(1)} = &-\frac{4\a_\pm^0}{\bm{\vec{k}}^4}\left(-3r_H^{2}\a_{\pm,0}\d\pi^t_{\hp{t}t}-\bm{\vec{k}}^2\d\pi_t + 4r_H^2 Q\a_{\pm,0}^2\d g^{(0)y}_{\hp{(0)y}y} - \frac{1}{2}Q\bm{\vec{k}}^2\d g^{(0)t}_{\hp{(0)t}t}\right)-\\
\label{n25} &- \frac{1}{2}Q\a_\pm^0\hat{z}_g - Q\hat{z}_a\, ,
\end{align}
where $\a_\pm^0\equiv \a_\pm(r=0)$ (defined in \eqref{E23b}), and we defined 
\begin{equation}
\label{n26} \hat{z}_a \equiv |\bm{\vec{k}}|(|\bm{\vec{k}}|\d A^{(0)}_t + \bm{\o}\d A^{(0)}_x)  
\end{equation}
\begin{equation}
\label{n27} \hat{z}_g = \left[ 2\bm{\o}|\bm{\vec{k}}|\d g^{(0)x}_{\hp{(0)x}t} + \bm{\o}^2\d g^{(0)x}_{\hp{(0)x}x} - \bm{\vec{k}}^2\d g^{(0)t}_{\hp{(0)t}t} + (\bm{\vec{k}}^2 - \bm{\o}^2)\d g^{(0)y}_{\hp{(0)y}y}  \right] \, .  
\end{equation}
Bold fonts indicate the quantities measured in units of the horizon radius $r_H$ (e.g. $\bm{\o}\equiv r_H \o$). Equations \eqref{n24}-\eqref{n25} imply that the helicity-0 correlators for $n=2$ cannot be linearly related to the the Kodama-Ishibashi correlators $G^\pm$. 

As shown in \cite{Edalati:2010pn}, the stress-tensor and current correlators are instead linearly related to a different set of master variables, given by
\begin{equation}
\label{n28} \Psi_\pm \equiv \frac{4r_H^3\a_\pm^0Q}{\bm{\vec{k}}^2}\Phi_\pm + f(r)r_H\pa_r\Phi_\pm \, .
\end{equation}
However, the Schr\"odinger potential associated with $\Psi_\pm$ takes the form 
\begin{equation}
\label{n29} V^\pm_\Psi(z) = V_\pm(z) - \frac{4Q\a_\pm^0}{\vec{k}^2}\frac{\g_\pm'(z)}{\g_\pm(z)} + \frac{\g_\pm''(z)}{2\g_\pm(z)} - \frac{(\g_\pm')^2}{4\g_\pm^2} \, ,
\end{equation}
with 
\begin{equation}
\label{n210} \g_\pm(z) = \o^2-V_\pm(z) + \frac{16Q^2(\a^0_\pm)^2}{\vec{k}^4} \, . 
\end{equation}
Since $V_\pm(z)$ is not always negative, the potential \eqref{n29} will generically have singularities at finite $z$, implying that the correlators $\Pi^\pm$ associated with $\Psi_\pm$ may admit zeroes. Note that, from the definition \eqref{n28}, one can infer that the $\Pi^\pm$ are related to the $G^\pm$ as \cite{Edalati:2010pn}
\begin{equation}
\label{n211} \Pi_R^\pm(\o,\vec{k}) = r_H^{-1}\frac{a(\vec{k})^2+\bm{\vec{k}}^2-\bm{\o}^2 + r_Ha(\vec{k})G^\pm_R(\o,\vec{k})}{a(\vec{k})+r_HG^\pm_R(\o,\vec{k})} \, , \quad a(\vec{k}) \equiv \frac{4r_H^3\a_\pm^0Q}{\bm{\vec{k}}^2} \, .
\end{equation}

\subsection{\texorpdfstring{$\bm{n\geq 4}$}{n≥4}}

We now discuss the case where the number of boundary spatial dimensions is larger than 4. In this case, the near-boundary expansion of the KI fields takes the form
\begin{equation}
\label{n41} \Phi_\pm(r) = \Phi_\pm^{(0)}r^{1-\n} (1+\OO(r^2)) + \Phi_\pm^{(1)}r^\n(1+\OO(r^2)) \, ,    
\end{equation}
with $\n=(n-2)/2\geq1$. For the metric and gauge field we have
\begin{equation}
\label{n42} \d g^\m_{\hp{\m}\n} = \d g^{(0)\m}_{\hp{(0)\m}\n}\left(1+r^2a_n(\o,\vec{k})+\OO(r^4)\right) + r^{n+1}\left(b_n(\o,\vec{k})\d g^{(0)\m}_{\hp{(0)\m}\n}\log{r}+\d\pi^{\m}_{\hp{\m}\n}(1+\OO(r^2))\right)\! ,
\end{equation}
\begin{equation}
\label{n43} \d A_\m = \d A_\m^{(0)}\left(1+ c_n(\o,\vec{k})r^2+\OO(r^4)\right) + r^{n-1}\left(d_n(\o,\vec{k})\d A_\m^{(0)}\log{r}+\d\pi_\m(1+\OO(r^2))\right) \, ,
\end{equation}
where $b_n$ and $d_n$ are 0 for even $n$. It can then be checked that $\Phi_\pm^{(0)}$ can be expressed in terms of the sources $\d g^{(0)\m}_{\hp{(0)\m}\n},\d A_\m^{(0)}$ only, without the vev's $\d\pi^{\m}_{\hp{\m}\n},\d\pi_\m$. In this case, there are therefore linear relations between the stress-tensor and current correlators, and the KI correlators $G^\pm$.  

\section{Poles of the stress-tensor correlators}

\label{Sec::AppE}

We present in this appendix the results of a numerical calculation of the low-lying poles of the helicity-0 stress-tensor correlators analyzed in section \ref{sec:sound}, for $n=3$ boundary spatial dimensions. The caculation was based on the Mathematica package of \cite{Jansen:2017oag}. As discussed in section \ref{sec:sound}, the poles split into two families: the sound channel associated with the $\Phi^+$ master variable, and the diffusive channel associated with $\Phi^-$.

\subsection{Diffusive channel}

\begin{figure}[h]
\begin{center}
\includegraphics[scale=1.]{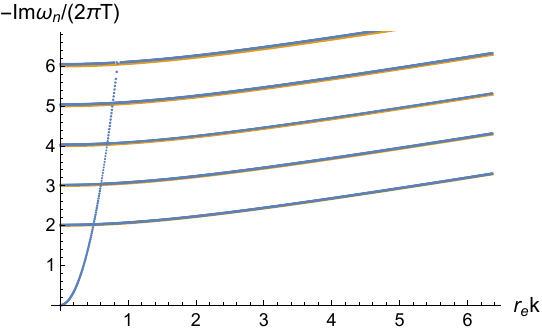}
\caption{Imaginary part of the low-lying poles of the $G^-$ retarded correlator, in units of $2\pi T$ and as a function of momentum $r_e|\vec{k}|$, for $n=3$ and $r_e T = (50\pi)^{-1}$. The blue dots are the numerical results whereas the orange lines show the poles of the IR CFT$_1$ correlator \eqref{dc1}.}
\label{Fig:poles_Gm}
\end{center}
\end{figure}

Figure \ref{Fig:poles_Gm} shows the low-lying poles of the diffusive channel (which are purely imaginary), for $r_e T = (50\pi)^{-1}\simeq6.4\times 10^{-3}$. The behavior is identical to what was found in \cite{Davison:2022vqh} for $n=2$: there is a single gapless pole which is diffusive $\o_D(\vec{k}) = -iD\vec{k}^2 + \OO(\vec{k}^4)$, and, up to $\OO(r_e T)$ corrections, the soft poles equal the IR poles (orange lines)
\begin{equation}
\label{dc1} \o_m^{\text{IR}}(\vec{k}) = -i2\pi T(\D_-(\vec{k})+m) \, ,
\end{equation}
with the IR dimension given by \eqref{E221}. 

\subsection{Sound channel}

\begin{figure}[h]
\begin{center}
\includegraphics[scale=1.]{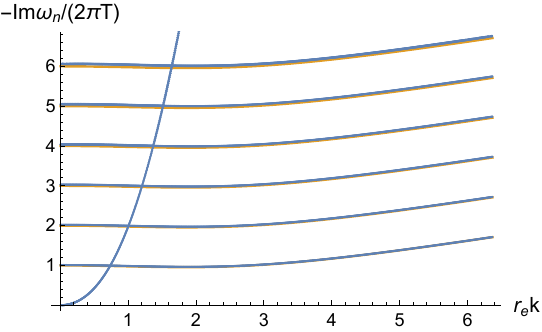}
\caption{Imaginary part of the low-lying poles of the $G^+$ retarded correlator in units of $2\pi T$ and as a function of momentum $r_e|\vec{k}|$, for $n=3$ and $r_e T = (50\pi)^{-1}$. The blue dots are the numerical results whereas the orange lines show the poles of the IR CFT$_1$ correlator \eqref{sc1}.}
\label{Fig:Impoles_Gp}
\end{center}
\end{figure}

\begin{figure}[h]
\begin{center}
\includegraphics[scale=1.]{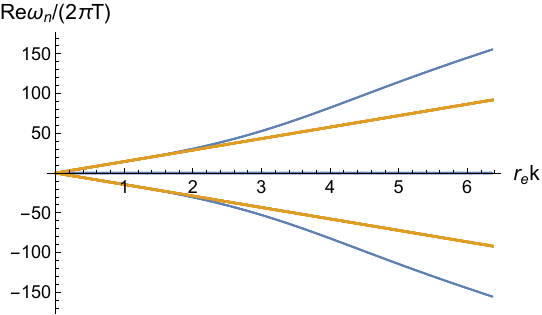}
\caption{Real part of the poles shown in figure \ref{Fig:Impoles_Gp}. Only the sound poles have finite real parts. The orange line shows the low momentum behavior $\o_{s,\pm}^0(\vec{k})=\pm c_s|\vec{k}| = \pm |\vec{k}|/\sqrt{n}$.}
\label{Fig:Repoles_Gp}
\end{center}
\end{figure}

The low-lying poles of the $G^+$ retarded correlator are shown in figures \ref{Fig:Impoles_Gp} (imaginary part) and \ref{Fig:Repoles_Gp} (real part), still for $n=3$ and $r_eT=(50\pi)^{-1}$. Two sound poles $\o_{s,\pm}(\vec{k}) = \pm c_s|\vec{k}| - i\G \vec{k}^2 + \OO(\vec{k}^4)$ are clearly identified, and the soft poles agree with the IR poles (orange lines in figure \ref{Fig:Impoles_Gp})
\begin{equation}
\label{sc1} \o_m^{\text{IR}}(\vec{k}) = -i2\pi T(\D_+(\vec{k})+m) \, ,
\end{equation}
up to $\OO(r_e T)$ corrections. Figure \ref{Fig:Repoles_Gp} also shows that the sound poles transition from hydrodynamic to relativistic behavior at large momentum, as has been observed previously for $n=2$ \cite{Edalati:2010pn}. 

\section{Scalar poles of the magnetic black brane}

\label{Sec::AppH}

\begin{figure}[h]
\begin{center}
\includegraphics[scale=1.]{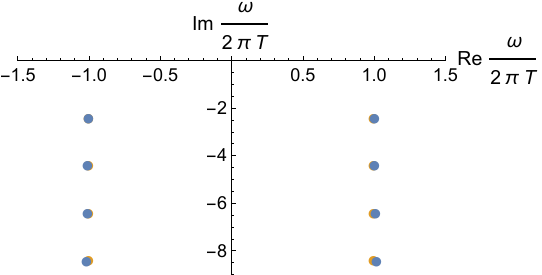}
\caption{First four poles of the retarded correlator for a massless scalar on top of the near-extremal magnetic black-brane background, for $r_e T \simeq 6.1\times 10^{-3}$, $k_x=2\pi T$ and $k_y^2=\sqrt{d-1}B$. The blue dots are the numerical results whereas the orange dots show the poles of the IR CFT$_2$ correlator \eqref{pmb}.}
\label{Fig:Cpoles_mb}
\end{center}
\end{figure}
\begin{figure}[h]
\begin{center}
\includegraphics[scale=1.]{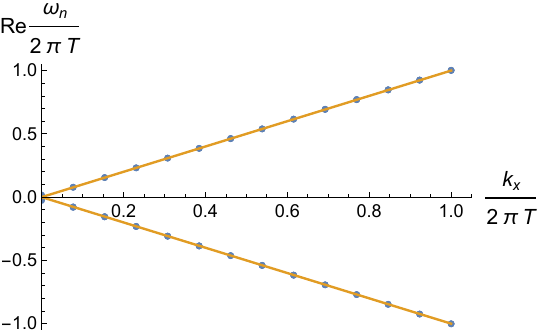}
\caption{Evolution of the real part of the poles of figure \ref{Fig:Cpoles_mb} for $k_y=0$ and $k_x$ between 0 and $2\pi T$. The blue dots are the numerical results, which are compared to the poles of the IR CFT$_2$ correlator \eqref{pmb} (orange lines).}
\label{Fig:Repoles_mb}
\end{center}
\end{figure}
\begin{figure}[h]
\begin{center}
\includegraphics[scale=1.]{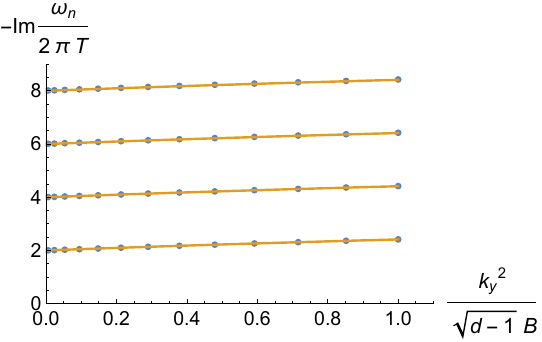}
\caption{Evolution of the imaginary part of the poles of figure \ref{Fig:Cpoles_mb} for $k_x=0$ and $k_y^2$ between 0 and $\sqrt{d-1}B$. The blue dots are the numerical results, which are compared to the poles of the IR CFT$_2$ correlator \eqref{pmb} (orange lines).}
\label{Fig:Impoles_mb}
\end{center}
\end{figure}

We present in this appendix the results of a numerical calculation for the low-lying poles of the scalar correlator on the near-extremal magnetic black-brane background discussed in section \ref{sec:mb}. In this case the background itself is numerical, for which the package of \cite{Jansen:2017oag} does not seem to handle higher working precision, which is necessary to compute more than the first two poles. Our calculation instead used a Mathematica script which is based on the same standard algorithm as \cite{Jansen:2017oag}, but allows to control the working precision throughout. 

Figures \ref{Fig:Cpoles_mb}, \ref{Fig:Repoles_mb} and \ref{Fig:Impoles_mb} present numerical results for the poles of a massless scalar at $r_e T \simeq 6.1\times 10^{-3}$, with $r_e\sim B^{-1/2}$ the extremal horizon radius. Figure \ref{Fig:Cpoles_mb} shows the poles in the complex plane for a fixed value of the momentum $k_x=2\pi T$ and $k_y^2=\sqrt{d-1}B$, whereas the momentum dependence is displayed in figures \ref{Fig:Repoles_mb} ($k_x$) and \ref{Fig:Impoles_mb} ($k_y$). These results indicate that the low-temperature scalar poles approach the IR poles (shown in orange) 
\begin{equation}
\label{pmb} \o_{n,\pm}^{\text{IR}}(k_x,k_y) = \pm k_x - i2\pi T(\D(k_y)+2n)  \, ,
\end{equation}
with $\D(k_y)$ given in \eqref{M21}. 

\newpage

\bibliographystyle{JHEP}
\bibliography{hpf}

\end{document}